\documentclass[11pt,authoryear,longnamesfirst]{elsarticle}

\let\today\relax
\makeatletter
\def\ps@pprintTitle{%
    \let\@oddhead\@empty
    \let\@evenhead\@empty
    \def\@oddfoot{\footnotesize\itshape
         {} \hfill\today}%
    \let\@evenfoot\@oddfoot
    }
\makeatother

\usepackage{amsmath, amssymb, amsthm, bbm}
\usepackage{booktabs}
\usepackage[top=2.5cm, left=3cm, right=3cm, bottom=2.5cm]{geometry} % Margins
\usepackage{multirow}
\usepackage[group-separator={,}]{siunitx}
\usepackage{hyperref} % HRef / Citations
\usepackage{soul} % Highlighting
\usepackage{xcolor} % Link / Citation colors
\usepackage{graphicx} % Images / resizebox

\newcommand\totalstocks{6{,}061}
\newcommand\fullstocks{1{,}445}
\newcommand\dowstocks{27}
\newcommand\nasdaqstocks{69}

\DeclareMathOperator*{\argmin}{arg\,min}

\definecolor{lightblue}{HTML}{398EB9}
\definecolor{darkblue}{HTML}{0000EE}
\definecolor{darkgreen}{HTML}{296836}
\newcommand\myshade{85}
\colorlet{mylinkcolor}{darkblue}
\colorlet{mycitecolor}{darkblue}
\colorlet{myurlcolor}{darkblue}
\hypersetup{
	linkcolor  = mylinkcolor!\myshade!black,
	citecolor  = mycitecolor!\myshade!black,
	urlcolor   = myurlcolor!\myshade!black,
	colorlinks = true,
}

\begin{document}

\begin{frontmatter}

\title{HARd to Beat: The Overlooked Impact of Rolling Windows in the Era of Machine Learning}

% % AUTHORS %%%%%%%%%%%%%%%%%%%%%%%%%%%%%%%%%%%%%%%%%%%%%%%%%%%%%%%%%%%%%%%%%%%%
% Leave this section commented out so that the paper is blinded for review.
% Group authors per affiliation:
\author[sg]{Francesco Audrino}
\address[sg]{University of St.Gallen, Switzerland}

\author[sg]{Jonathan Chassot\corref{cor}}
% \address[jc]{University of St.Gallen, Switzerland}

% Only give the email address of the corresponding author
\cortext[cor]{Corresponding author}
\ead{jonathan.chassot@unisg.ch}
% %%%%%%%%%%%%%%%%%%%%%%%%%%%%%%%%%%%%%%%%%%%%%%%%%%%%%%%%%%%%%%%%%%%%%%%%%%%%%%%

% \maketitle
\begin{abstract}
	We investigate the predictive abilities of the heterogeneous autoregressive (HAR) model compared to machine learning (ML) techniques across an unprecedented dataset of {\fullstocks} stocks.
Our analysis focuses on the role of fitting schemes, particularly the training window and re-estimation frequency, in determining the HAR model's performance.
Despite extensive hyperparameter tuning, ML models fail to surpass the linear benchmark set by HAR when utilizing a refined fitting approach for the latter.
Moreover, the simplicity of HAR allows for an interpretable model with drastically lower computational costs.
We assess performance using QLIKE, MSE, and realized utility metrics, finding that HAR consistently outperforms its ML counterparts when both rely solely on realized volatility and VIX as predictors.
Our results underscore the importance of a correctly specified fitting scheme. 
They suggest that properly fitted HAR models provide superior forecasting accuracy, establishing robust guidelines for their practical application and use as a benchmark.
This study not only reaffirms the efficacy of the HAR model but also provides a critical perspective on the practical limitations of ML approaches in realized volatility forecasting.
\end{abstract}

\begin{keyword}
    Forecasting practice, HAR, Machine learning, Realized volatility, Volatility forecasting
% Suggested keywords are listed at https://ijf.forecasters.org/keywords/
\end{keyword}

\end{frontmatter}
\clearpage
\section{Introduction}\label{sec:introduction}
Accurate forecasts of realized volatility are essential for various financial applications, including risk management, derivative pricing, and portfolio optimization.
The modeling of financial volatility has a deep-rooted history in academic literature and has undergone extensive exploration in recent decades \citep[see, e.g.,][for comprehensive reviews]{Bauwens2012,Takahashi2023}.

One simple model that stands out for its popularity and efficacy in realized volatility forecasting is the heterogeneous autoregressive (HAR) model, introduced by \citet{Corsi2009}.
This model predicts the next day's realized volatility by calculating a weighted average based on the realized volatilities of the preceding day, week, and month.
Despite its simplicity, the HAR model has established itself as a widely adopted benchmark in the field, consistently demonstrating strong forecasting accuracy and often outperforming more complex models.

The rise of machine learning (ML) techniques over the past years has marked a paradigm shift in financial econometrics, offering a new set of tools to tackle the challenges of volatility prediction.
This trend is supported by a growing body of literature, including studies by \citet{Audrino2016,Audrino2020,Bucci2020,Christensen2023,Zhang2023,Zhu2023}, among others, which highlight the diverse applications and potential benefits of ML in this area.

Given financial data's complex and often nonlinear nature, ML appears well-suited to uncover complex patterns that may elude traditional linear models such as the HAR model.
This intuition is supported by empirical evidence, which suggests that ML methods can indeed improve upon the forecasting accuracy of the HAR model \citep[see][for a recent survey]{Gunnarsson2024}.

\begin{figure}[!htb]
    \centering
    \includegraphics[width=\textwidth]{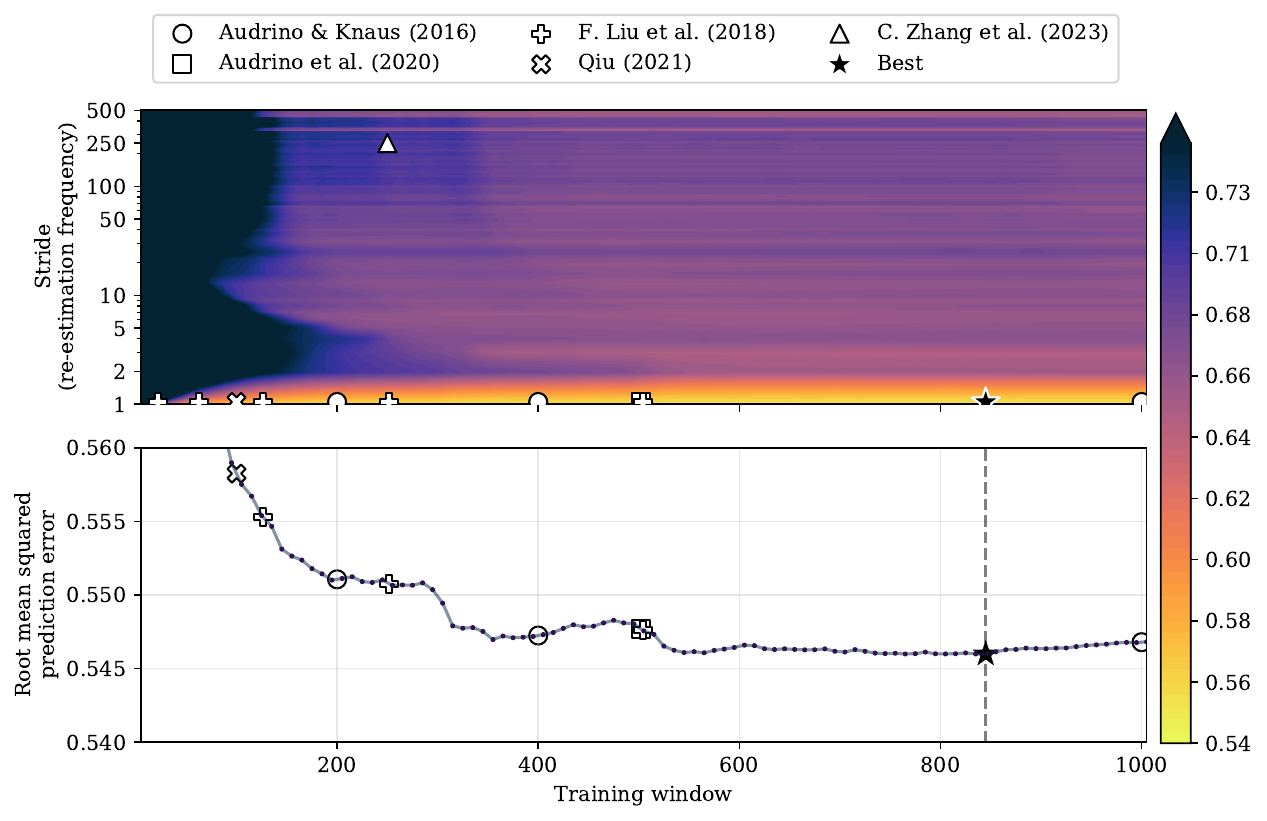}
    \caption{Top: Heatmap displaying the average root mean squared error (RMSE) of a HAR model with different fitting schemes for {\fullstocks} assets from January 2016 to December 2021. The $x$-axis shows the training window size, while the $y$-axis denotes the stride or re-estimation frequency. Color intensity varies with RMSE levels; darker colors indicate higher prediction errors. The fitting schemes used in previous studies are marked with distinct symbols for comparison. \\
    Bottom: Focused analysis on the scenario where the stride is $1$ (which emerges as the best-performing stride in the top plot). This plot provides a detailed look at how the size of the training window affects the RMSE, illustrating how larger training windows generally lead to lower prediction errors.}
    \label{fig:har_heatmap}
\end{figure}

In this study, we argue that the reality is more subtle and, despite their appeal, ML methods are not a panacea for volatility forecasting.
Specifically, our findings demonstrate that the effectiveness of the HAR model is highly sensitive to the choice of fitting scheme.
HAR models are typically estimated in a rolling window fashion, where the model is re-estimated at regular intervals using a fixed (or increasing) number of past observations.
As such, a researcher must specify two key parameters when estimating the HAR model: the training window size and the re-estimation frequency.
As Figure~\ref{fig:har_heatmap} illustrates, the specification of these parameters profoundly impacts the model's forecasting accuracy.
This study shows that the optimal fitting schemes are exceptionally hard to beat, even when using sophisticated ML techniques.
The most notable feature of this figure is the substantial deterioration of the forecasting performance when the model is not re-estimated daily, even when the re-estimation frequency is relatively high, e.g., every two or five days.
While the impact of the training window size is less pronounced than that of the re-estimation frequency, the bottom half of Figure~\ref{fig:har_heatmap} shows that this parameter also plays a crucial role in obtaining accurate forecasts.
In particular, larger training windows generally lead to lower prediction errors, with the optimal window size being roughly two and a half to four years.
Naturally, a higher re-estimation frequency and larger training windows leads to higher computation costs.
However, this computational burden is greatly alleviated by the simplicity and linearity of the HAR model, which allows for efficient estimation even with large datasets.

\section{Related Literature}\label{sec:related_work}
Realized volatility forecasts have been the subject of extensive research in finance, and numerous econometric models have been proposed over the years \citep{Bauwens2012}.
In recent years, ML techniques have gained popularity in financial econometrics, offering a new set of tools to tackle the challenges of volatility prediction.
\citet{Gunnarsson2024} provide an extensive modern survey of volatility forecasts with ML.
Our work focuses on literature that applies ML techniques to RV forecasts, specifically emphasizing studies that use the HAR model as a benchmark.
We summarize the key findings of these studies below and provide an overview of the different fitting schemes these studies use for the baseline HAR model in Table~\ref{tbl:studies}.

\citet{Audrino2016} apply the least absolute shrinkage and selection operator (lasso) to the task of RV forecasting.
They show that if HAR is the true data-generating process, lasso can recover the lag structure of the HAR model asymptotically. 
Empirically, they find that the lasso and the HAR model suggest a different lag structure, and the two models perform similarly in terms of out-of-sample forecasting accuracy.
\citet{Audrino2020} also apply the lasso methodology to RV forecasting.
However, they go beyond adding simple lags and extend the baseline model with covariates derived from economic variables and investor attention and sentiment measures.
Their findings suggest that including these additional features can improve forecast accuracy compared to the HAR model.
\citet{Liu2018} propose RV forecasts using recurrent neural networks (RNNs).
They find evidence suggesting that RNNs can improve forecast accuracy when the training sample is small.
However, as the training sample size increases, the HAR model outperforms the RNNs. 
\citet{Qiu2021} proposes an ML model based on support vector regression and finds that it outperforms the HAR model in an out-of-sample forecasting exercise on Bitcoin data.
\citet{Zhang2023} compare the HAR model with a wide range of ML models, including lasso, random forests (RFs), gradient-boosted trees (GBTs), feedforward neural networks (FFNNs), and long short-term memory networks (LSTMs).
They consider several training specifications, including individual stock estimation and pooled estimation.
They find that augmenting the features with intraday data can improve forecast accuracy when using ML models, particularly FFNNs.
\citet{Christensen2023} compare an extensive collection of ML models with various HAR model specifications, first using the information set of the original HAR model and then augmenting it with firm characteristics and macroeconomic variables, where the richer information set may allow ML models to demonstrate superior performance.
While they use a rolling window scheme for the HAR model, they do not mention the exact training window and stride length used.
Thus, we cannot report this information in Figure~\ref{fig:har_heatmap} or Table~\ref{tbl:studies}.

Notably, when considering Figure~\ref{fig:har_heatmap} and Table~\ref{tbl:studies} in conjunction with the findings of the studies summarized above, we observe that studies with a daily re-estimation and long training windows tend to find that the HAR model is competitive with or outperforms ML models.
In contrast, studies with suboptimal fitting schemes for the HAR model tend to find that ML models outperform the HAR model.
As we demonstrate in this study, these discrepancies can easily be explained by the sensitivity of the HAR model's performance to the choice of fitting scheme, i.e., training window length and re-estimation frequency.

\begin{table}[h!]
    \begin{center}
    % \resizebox{\columnwidth}{!}{
    \begin{tabular}{l ccccc}
    \toprule
    \textbf{Study} & \textbf{Window Style} & \textbf{Training Window} & \textbf{Stride} & \textbf{Assets}\\
    \midrule
    \citet{Audrino2016}  & rolling & $200, 400, 1000, 2000$ & $1$ & $9$ \\
    \citet{Audrino2020} & rolling & $502$ & $1$ & $19$ \\
    \citet{Christensen2023} & rolling & --- & --- & $29$ \\
    \citet{Liu2018} & rolling & $22,63,126,252,504$ & $1$ & $4$ \\
    \citet{Qiu2021} & rolling & $100$ & $1$ & $1$ \\
    \citet{Zhang2023}  & expanding & $250$ & $250$ & $100$ \\
    This work & rolling & $630$ & $1$ & {\fullstocks}\\
    \bottomrule
    \end{tabular}%}
    \caption{Fitting schemes of HAR baselines used in different studies.}
    \label{tbl:studies}
    \end{center}
\end{table}

Lastly, our study does not extend the information set beyond the past RV series and the volatility index (VIX), the latter being a typical and cost-effective addition for improving the forecast accuracy of HAR models \citep{Audrino2020,Buncic2016,Zhang2020}.
While other studies, such as \citet{Audrino2020,Christensen2023}, demonstrate that extending the information set can improve forecasts, our research concentrates on optimizing the predictive power of the HAR model using a narrow and well-established information set that any practitioner can easily implement.
This focus allows us to assess the intrinsic forecasting power of the HAR model and the extent to which ML models can improve upon it, leaving the exploration of richer information sets for future research.
\section{Data}\label{sec:data}
\subsection{Realized Volatility}\label{subsec:realized_volatility}
We consider a financial asset whose price evolves following a stochastic model represented by $$dp(t) = \mu(t) dt + \sigma(t) dW(t),$$ in which $p(t)$ denotes the logarithm of the price at any given time $t$, $\mu(t)$ embodies the trend or drift, $\sigma(t)$ is the level of volatility, and $W(t)$ models a standard Brownian motion. 
Within this framework, a theoretical concept for measuring volatility over a specific period $(t-h, t]$ emerges as the integrated variance (IV), mathematically expressed as $$\mathrm{IV}(t, h) = \int_{t-h}^{t} \sigma^2(s) ds,$$ where $h$ represents the duration or lookback period. 
However, the theoretical construct of IV is not directly observable in real-world scenarios, necessitating estimations. 
The concept of realized volatility (RV), introduced through foundational contributions by \citet{Andersen2001,Andersen2003,Barndorff-Nielsen2002,Barndorff-Nielsen2002a} has become a widely adopted metric for estimating the IV in practice. 
This measure leverages high-frequency intraday data to provide a consistent estimator of the IV.
In line with previous research \citep[e.g.,][]{Andersen2003,Zhang2023}, we adopt the logarithmic RV estimator, which is well-known to outperform its \textit{standard} counterpart \citep{Clements2021,Taylor2017}. We define the one-day logarithmic RV as \begin{equation}\label{eq:rv}
    \mathrm{RV}_t = \log\left( \sum_{i=0}^{m-1} r_{t- \Delta i }^2\right),
\end{equation} where $r_{t- \Delta i }$ denotes the intraday logarithmic return over $\Delta$ minutes, and $m$ represents the number of intervals in a trading day\footnote{While the primary focus of this study is on the fitting schemes and performance of the HAR model, it is important to note that our analysis does not specifically address the complications arising from jumps or microstructure noise in the financial data. When such factors are present, our approach estimates the total IV contributions, including jumps.}.
In particular, we set $\Delta = 5$, corresponding to an interval widely used in the literature and shown to be difficult to outperform \citep{Liu2015}.

\subsection{Intraday Data}\label{subsec:intraday_data}
In our analysis, we conduct a thorough examination of stocks traded on U.S. exchanges. 
Specifically, we focus on the constituents of the Center for Research in Security Prices' (CRSP) U.S. Stock Database, for which we can obtain matching ticker symbols in the New York Stock Exchange (NYSE) Trade and Quote (TAQ) database. 
We source high-frequency intraday trade data for all selected stocks from the TAQ database from January 1, 2015, to October 31, 2023. 
These data are processed following the methodology suggested by \citet{Barndorff-Nielsen2009}.

We query high-frequency data for a total of {\totalstocks} stocks. 
Among these, {\fullstocks} stocks remain consistently present throughout the nine years. Out of these {\fullstocks} stocks, {\dowstocks} belong to the DJIA index (out of the 30 total constituents), and {\nasdaqstocks} are part of the Nasdaq-100 index (out of the 100 total constituents).
Our study focuses on the entire set of {\fullstocks} stocks to provide a comprehensive and unprecedented analysis of the performance of the models across a broad spectrum of the market.
However, we also perform an additional analysis of the DJIA and Nasdaq-100 stocks in the appendix to validate our main results further.
These specific subsets of stocks are selected for their economic significance and increased liquidity and trading activity compared to the entire range of stocks, making the results obtained in the appendix potentially more relevant to practical applications.
In so doing, we aim to demonstrate that our main findings hold even in these more focused and economically relevant universes, which are more often studied in financial literature. 
Thus, we reinforce the applicability and robustness of our results.
This starkly contrasts our work with previous research, which typically focuses on a rather limited number of stocks (see Table~\ref{tbl:studies}).
\section{Methodology}\label{sec:models}

\subsection{Heterogeneous Autoregressive Model}\label{sec:har}
The HAR model, introduced by \citet{Corsi2009}, provides a parsimonious and economically interpretable linear model for RV forecasting that is straightforward to implement and offers a high degree of flexibility and adaptability, as evidenced by several extensions and modifications proposed in the literature \citep{Corsi2012}.

In its canonical formulation, the HAR model is expressed as \begin{equation}\label{eq:har}
    \mathrm{RV}_{i,t+1}^{(d)} = c + \beta^{(d)} \mathrm{RV}_{i,t}^{(d)} + \beta^{(w)} \mathrm{RV}_{i,t}^{(w)} + \beta^{(m)} \mathrm{RV}_{i,t}^{(m)} + \varepsilon_{i,t+1},
\end{equation} where $\mathrm{RV}_{i,t}^{(d)}$ denotes the logarithm of the daily realized volatility \eqref{eq:rv} of stock $i$ at time $t$, $\mathrm{RV}_{i,t}^{(w)} = \frac{1}{5}\sum_{h=1}^5 \mathrm{RV}_{i,t-h+1}^{(d)}$ and $\mathrm{RV}_{i,t}^{(m)} =  \frac{1}{22}\sum_{h=1}^{22} \mathrm{RV}_{i,t-h+1}^{(d)}$ represent the weekly and monthly realized volatilities, respectively, and $\{\varepsilon_{i,t}\}_{t\in\mathbb{Z}}$ is a zero-mean innovation process.

An especially compelling feature of the HAR model is its linearity, allowing for the effortless incorporation of additional predictors at a low computational cost.
In particular, the Chicago Board Options Exchange (CBOE) Volatility Index (VIX) has been demonstrated to markedly improve the model's predictive capability when included as an additional predictor \citep{Audrino2020,Buncic2016,Zhang2020}.
The VIX measures the stock market's expectation of future volatility. 
It is calculated using the implied volatilities of a wide range of S\&P 500 index options and thus provides additional market-wide information not captured by an individual stock's realized volatility.
The resulting model is referred to as HAR-VIX and is expressed as \begin{equation}\label{eq:har_vix}
    \mathrm{RV}_{i,t+1}^{(d)} = c + \beta^{(d)} \mathrm{RV}_{i,t}^{(d)} + \beta^{(w)} \mathrm{RV}_{i,t}^{(w)} + \beta^{(m)} \mathrm{RV}_{i,t}^{(m)} + \beta^{(v)} \mathrm{VIX}_{t} + \varepsilon_{i,t+1},
\end{equation} where $\mathrm{VIX}_{t}$ denotes the VIX at time $t$.

The HAR and HAR-VIX models are typically estimated using ordinary least squares (OLS) regression per stock.
However, \citet{Clements2021} suggest that alternative estimation techniques, such as weighted least squares (WLS) and robust regression, can significantly enhance forecast accuracy.
Furthermore, pooled approaches have garnered attention, wherein the HAR model is simultaneously estimated across various stocks. This method capitalizes on the entire dataset to enhance forecast accuracy, as suggested by \citet{Patton2015,Zhang2023,Bollerslev2018}.

To conduct a comprehensive comparison between the HAR and HAR-VIX models and their machine learning counterparts, we estimate equations \eqref{eq:har} and \eqref{eq:har_vix} using both OLS and WLS regression methods.
These estimations are performed individually for each stock and in a pooled manner, resulting in eight distinct specifications. 
The HAR models are estimated using a rolling window approach described in Section~\ref{sec:estimation}.

\subsection{Machine Learning Models}\label{sec:ml_models}
\subsubsection{Least Absolute Shrinkage and Selection Operator}\label{sec:lasso}
The Least Absolute Shrinkage and Selection Operator (lasso) introduced by \citet{Tibshirani1996} is a linear regression method that applies a penalty on the absolute values of the regression coefficients.
This penalty shrinks the coefficients of less important predictors to zero, effectively performing variable selection.
Lasso is especially attractive in high-dimensional settings, offering models that are both interpretable and predictively proficient \citep[see, e.g.,][for a textbook treatment]{Hastie2015,Hastie2009}.
Consequently, the lasso emerges as an appealing ML candidate for RV forecasting, with the potential to provide an economically tractable model.
\citet{Audrino2016,Audrino2020,Zhang2023,Christensen2023} have previously applied the lasso to the problem of RV forecasting.
Lasso modifies the standard linear regression model by solving the following optimization problem: \begin{equation}\label{eq:lasso}
    \hat{\boldsymbol{\beta}} = \argmin_{\boldsymbol{\beta}} \left\{ \frac{1}{2n} \left\| \mathbf{y} - \mathbf{X}\boldsymbol{\beta} \right\|_2^2 + \lambda \left\| \boldsymbol{\beta} \right\|_1 \right\},
\end{equation} where $\mathbf{y}$ is the vector of outcomes, $\mathbf{X}$ is the design matrix, $\boldsymbol{\beta}$ is the vector of regression coefficients, and $\lambda$ is a penalty parameter that controls the amount of shrinkage applied to the coefficients.
% The lasso thus has a hyperparameter $\lambda$ that needs to be tuned.
% In our analysis, we use cross-validation to select the optimal value of $\lambda$ for each stock and each model specification.

\subsubsection{Random Forest}\label{sec:rf}
The Random Forest (RF) algorithm, proposed by \citet{Breiman2001}, is a popular ensemble learning method that constructs numerous decision trees and aggregates their predictions to enhance predictive accuracy and control overfitting.
RFs operate as bagging algorithms, as described by \citet{Breiman1996}; i.e., they leverage bootstrap sampling to create multiple predictors.
This technique reduces variance by averaging the predictions from various low-bias, high-variance models, enhancing the model's generalization error.

For a comprehensive understanding of RF, readers are directed to seminal works by \citet{Breiman2001} and \citet{Hastie2009}.
Here, we only provide a concise summary of the RF construction process:
\begin{enumerate} \setlength\itemsep{0em}
    \item \textbf{Bootstrap Sampling}: Draw a bootstrap sample from the original dataset.
    \item \textbf{Tree Construction}: Individually grow each tree by randomly selecting a subset of features at each decision split, effectively de-correlating the trees.
    \item \textbf{Forest Creation}: Repeat the first two steps to generate many trees, forming a \textit{forest}.
\end{enumerate}
The final model prediction is derived by averaging the predicted outcomes across all the trees, thereby mitigating any individual tree's prediction errors.

In the context of RV forecasting, \citet{Luong2018} have found RFs to outperform the performance of the traditional HAR model. 
These results are somewhat mitigated by \citet{Christensen2023}, who find that RFs only outperform the HAR model when the information set is expanded to include firm characteristics and macroeconomic variables as additional features.
However, \citet{Christensen2023} do not perform any hyperparameter tuning for their RF model, which may negatively impact its performance.
Moreover, when the information set of the HAR model is expanded to include the same features as the RF model, the performance discrepancy is no longer statistically significant.

\subsubsection{Gradient Boosted Trees}\label{sec:gbt}
Gradient Boosted Trees (GBTs), introduced by \citet{Friedman2001}, is another powerful ensemble learning technique that sequentially builds decision trees, unlike RFs, which construct trees independently.
Each tree in a GBT model is constructed to correct the residual errors of previous trees, thereby enhancing the model's predictive power with each iteration.

An in-depth discussion of GBTs is provided in the original paper by \citet{Friedman2001} and the comprehensive work of \citet{Hastie2009}.
We restrict our discussion to a brief overview of the GBT construction process:
\begin{enumerate} \setlength\itemsep{0em}
    \item \textbf{Initialize a Base Model}: Start with a simple predictive model (e.g., constant prediction of the mean).
    \item \textbf{Sequential Tree Building}: For each subsequent tree, focus on correcting the errors made by previous trees. This is achieved by fitting a new tree to the residuals of the current model.
    \item \textbf{Combine Predictors}: Aggregate the predictions from all individual trees using a weighted sum to obtain the final model prediction, where the learning rate determines the weights. This hyperparameter controls the contribution of each tree.
\end{enumerate}

In the context of RV forecasting, \citet{Zhang2023} do not find gradient boosting to outperform the HAR model or other ML models.
Similarly, \citet{Christensen2023} find that GBTs do not outperform the HAR model when the information set is limited to RV series.
On the other hand, if the information set is expanded to include firm characteristics and macroeconomic variables, they find that GBTs outperform the HAR model.
However, as with the RF model, the HAR model's performance is not statistically different from the GBT model when the information set is expanded to include the same features as the GBT model.

\subsubsection{Feedforward Neural Network}\label{sec:ffnn}
Feedforward Neural Networks (FFNNs), often referred to as Multilayer Perceptrons (MLPs), is a foundational class of artificial neural networks composed of multiple layers of neurons, with each neuron in a layer connected to all neurons in the subsequent layer via weighted connections.
A typical FFNN consists of an input layer, one or more hidden layers, and an output layer.

The primary appeal of FFNNs lies in their ability to approximate virtually any function, given a sufficient number of neurons and layers \citep{Hornik1991}.
This ability makes them highly versatile and well-suited for capturing complex patterns in high-dimensional data; as such, FFNNs have become a standard tool in the ML toolkit.
Several textbook references provide a comprehensive overview of FFNNs, including \citet{Goodfellow2016} and \citet{Hastie2009}.

Regarding realized volatility, \citet{Zhang2023} find that FFNNs outperform the HAR model for intraday RV forecasting.
However, these results do not hold when it comes to daily volatility, for which the HAR model is designed.
On the other hand, \citet{Christensen2023} find that several FFNN architectures outperform the HAR model even when the information set is limited to be the same as what the HAR model uses.

\subsection{Estimation Strategy}\label{sec:estimation}
One of the most notable strengths of ML techniques is their ability to process high-dimensional data, which facilitates the inclusion of diverse covariates beyond the handcrafted features of the HAR models.
As discussed in Section~\ref{sec:related_work}, this study deliberately focuses on the plain HAR model setting without extending the information set to include predictors outside the RV series and the VIX.
Nonetheless, we augment the information set by incorporating 100 lags of the RV series, allowing the ML models to potentially capture long-term dependencies in the data that may elude the HAR models.
Enlarging the information set by including additional features, such as sentiment indicators in the spirit of \citet{Audrino2020} or firm characteristics as in \citet{Christensen2023}, is left for future research.

ML approaches often require fine-tuning hyperparameters, and the particular choice of hyperparameters significantly impacts the model's predictive performance.
The methods outlined in Section~\ref{sec:ml_models} are no exception to this rule.
However, the computational burden of hyperparameter optimization presents a substantial obstacle for the rolling re-estimation of ML models.
Therefore, implementing a rolling window methodology with an extensive training period and minimal re-estimation frequency proves impractical for analyzing a wide array of assets.
Table~\ref{tbl:runtimes} summarizes the fitting times for the HAR and ML models for a single stock.
The rolling window approach used for the HAR models necessitates re-estimating the model for each observation in the test set, resulting in a total of 460 re-estimations.
In contrast, the fitting times for the ML models represent the average time required using a static training window and hyperparameter tuning on the validation set.
Estimating the ML models using a rolling window approach would require repeating this process for each new observation in the test set, resulting in a substantial increase in computational time (roughly 460 times longer than the static training window approach).
The goal of Table~\ref{tbl:runtimes} is not to provide a detailed comparison of the computational efficiency of the models but to illustrate the computational challenges associated with the rolling window approach for ML models.
Moreover, the table highlights the runtime differences for our selection of hyperparameter candidates, which are driven by the fitting times of the models (i.e., we choose the optimal $\lambda$ for the lasso model using $1{,}000$ candidates as it is an ML model with a relatively short fitting time, while we use only $27$ candidates for the RF model, which results in a relatively shorter fitting time).

\begin{table}[h!]
    \begin{center}
    \begin{tabular}{l rr}
    \toprule
    \textbf{Model} & \multicolumn{2}{c}{\textbf{Runtime}}\\
    & \multicolumn{1}{c}{\textbf{No VIX}} & \multicolumn{1}{c}{\textbf{VIX}}\\
    \midrule
    HAR (OLS) & 1.77 & 1.78 \\
    HAR (WLS) & 1.86 & 1.88 \\
    lasso & 12.88 & 13.57 \\
    Random Forest & 13.16 & 13.13 \\
    Gradient Boosted Trees & 5.20 & 5.21 \\
    Feedforward Neural Network & 24.53 & 25.00 \\
    \bottomrule
    \end{tabular}
    \caption{Runtime (in seconds) for the models to fit on a single asset (AAPL). The HAR models are estimated using the rolling window approach; i.e., the models are re-estimated for each new observation in the test set, yielding a total of 460 re-estimations. In contrast, the ML models are estimated once for each hyperparameter configuration on the training set and validated in the validation set. The runtime is the average of 100 runs.}
    \label{tbl:runtimes}
    \end{center}
\end{table} 

Recent studies exploring multiple ML models for RV prediction across numerous stocks have adopted different strategies to address this computational challenge.
For instance, \citet{Christensen2023} tackle this issue by employing a static training window, dividing the dataset into 70\% training, 10\% validation, and 20\% test data.
Conversely, \citet{Zhang2023} opt for an expanding window method with an extended stride length, updating their models only every 250 days.
Notably, both studies find that ML models outperform the HAR model regarding forecast accuracy.
However, \citet{Zhang2023} benchmark their models against the HAR model with the same 250-day stride length, severely hampering the HAR model's performance without a substantial gain in computational efficiency for this simple linear model.
This approach is particularly disadvantageous for the HAR model, as the model's performance is highly sensitive to the choice of the re-estimation window, as illustrated in Figure~\ref{fig:har_heatmap}.\footnote{While \citet{Christensen2023} also use a rolling window for the HAR model, we cannot make an exact statement about the fitting scheme used, as they do not provide a detailed description of their training window and stride length for the HAR models.}
While Figure~\ref{fig:har_heatmap} illustrates the sensitivity of the HAR model's performance to the re-estimation frequency and training window size, it does so for HAR models estimated using a rolling window approach.
In contrast, the HAR models in \citet{Zhang2023} are estimated using an expanding window method.
We report the analogous heatmap for the expanding window method in the appendix in Figure~\ref{fig:har_heatmap_expanding}, which shows that the starting training window size has a smaller impact on the model's performance compared to the rolling window approach.
However, the re-estimation frequency remains a crucial determinant of the model's forecasting accuracy, even when using an expanding window method.

Adopting a strategy akin to that of \citet{Christensen2023}, we implement a static training window for our ML algorithms.
Specifically, our dataset is divided into a training period from January 2016 to December 2020, a validation period from January 2021 to December 2021, and a test period from January 2022 to November 2023, resulting in approximate splits of 64\%, 13\%, and 23\% for the training, validation, and testing sets, respectively.
In contrast, we opt for a rolling window method with a training duration of 630 days (approximately three and a half years) and daily stride length for the HAR and HAR-VIX models.
This choice is motivated by the findings in Figure~\ref{fig:har_heatmap}, indicating that the HAR model's performance is highly sensitive to the stride length. 
However, it remains relatively stable for training window lengths between two and four years when the stride length is set to one.

For each ML model, hyperparameter tuning is conducted individually for each stock on the validation set, using the candidate values outlined in \ref{sec:hyperparameters}.
While the set of assets we consider is far more extensive than used in any previous study, putting a further strain on computational resources required by the ML models, we ensure these candidates span a wide enough spectrum to provide comparable results to the hyperparameter choices studied in previous research \citep{Zhang2023,Christensen2023}.\footnote{Note that \citet{Zhang2023} only report the chosen hyperparameter values, not the candidates. Hence, we can only ensure these final values are a subset of our candidates.}
However, our primary aim is not to focus on exhaustive hyperparameter optimization but to demonstrate that out-of-the-box ML models do not effortlessly surpass a correctly fitted HAR or HAR-VIX baseline model, as previously suggested.
We provide a detailed discussion of the hyperparameters considered for each model in \ref{sec:hyperparameters}.

\subsection{Evaluation Metrics}\label{sec:metrics}
We evaluate the models' out-of-sample performance using the test set from January 2022 to November 2023.
We report the aggregated results across all {\fullstocks} stocks in the main text and provide additional insights into the performance of the models for the DJIA and Nasdaq-100 stocks in the appendix.
Although the DJIA and Nasdaq-100 stocks are less representative of the entire market, they provide a more focused view of economically significant stocks, and the current literature on RV forecasting often focuses solely on constituents of such indices \citep[e.g.,][]{Audrino2020,Christensen2023,Zhang2023}.
This justifies the inclusion of these subsets in our analysis, as they allow for a more direct comparison with previous research. 
This ensures that our results are robust and not only driven by including less liquid or less economically significant stocks.

\subsubsection{Statistical Loss Functions}\label{sec:statistical_metrics}
To assess the models' predictive accuracy, we use two standard evaluation metrics from the literature, namely the MSE, 
\begin{equation}
    \mathrm{MSE}_i = \frac{1}{T}\sum_{t=1}^T \left(\mathrm{RV}_{i,t}^{(d)} - \widehat{\mathrm{RV}}_{i,t}^{(d)}\right)^2,
\end{equation}
and the ``QLIKE'' loss function \citep[see][Eq. (24)]{Patton2011},
\begin{equation}
    \mathrm{QLIKE}_i = \frac{1}{T} \sum_{t=1}^T \left(\frac{\exp\left(\mathrm{RV}_{i,t}^{(d)}\right)}{\exp\left(\widehat{\mathrm{RV}}_{i,t}^{(d)}\right)} - \left(\mathrm{RV}_{i,t}^{(d)} - \widehat{\mathrm{RV}}_{i,t}^{(d)}\right) - 1 \right),
\end{equation}
where $T$ is the number of trading days in the test period, $\mathrm{RV}_{i,t}^{(d)}$ is the daily realized volatility of stock $i$ at time $t$, and $\widehat{\mathrm{RV}}_{i,t}^{(d)}$ is the model's forecast.

For each asset, we use the model confidence set (MCS) procedure \citep{Hansen2011} to compare the models' performance.
The MCS is a statistical procedure that compares multiple models based on their out-of-sample loss.
Due to its widespread use in the RV forecasting literature \citep[e.g.,][]{Audrino2020,Cipollini2021,Zhang2023,Christensen2023}, we do not provide a detailed explanation of the MCS procedure here and refer the reader to the original paper by \citet{Hansen2011} for a comprehensive description.
We conduct the MCS procedure at a 95\% confidence level for each stock individually, using the MSE and QLIKE loss functions as the metrics of interest.
Finally, we report the percentage of stocks for which each model is included in the collection of ``best'' models returned by the MCS procedure, providing a comprehensive overview of the models' performance across the entire set of stocks.

\subsubsection{Economic Performance}\label{sec:economic_metrics}
While MSE and QLIKE metrics measure statistical loss performance, they lack a direct economic interpretation. 
To address this shortcoming, \citet{Bollerslev2018} introduce a utility-based framework to assess risk models more comprehensively. 
This framework evaluates risk models through the lens of the expected utility of an investor with mean-variance preferences trading an asset with a constant Sharpe ratio. 
Investors aim to maintain a constant level of volatility by adjusting their portfolio position based on the models' forecasts, thereby creating a direct link between the models' predictions and the investors' utility.
We refer the reader to \citet{Bollerslev2018,Zhang2023} for a detailed explanation of the realized utility framework and note that the realized utility may be computed using the forecasts as 
\begin{equation}\label{eq:realized_utility}
    \mathrm{RU}_t = \frac{\mathrm{SR}^2}{\gamma} \cdot\left( \sqrt{\frac{\exp(\mathrm{RV}_{t+1})}{\exp(\widehat{\mathrm{RV}}_{t+1})}} - \frac{1}{2} \cdot \frac{\exp(\mathrm{RV}_{t+1})}{\exp(\widehat{\mathrm{RV}}_{t+1})}\right),
\end{equation}
where $\mathrm{SR}$ is the Sharpe ratio, $\gamma$ is the risk aversion parameter, $\mathrm{RV}_{t+1}$ is the realized volatility, and $\widehat{\mathrm{RV}}_{t+1}$ is the model's forecast.

For the sake of comparison, we implement the realized utility framework using the same specifications as \citet{Bollerslev2018,Zhang2023}, i.e., a risk aversion parameter of $\gamma = 2$ and a Sharpe ratio of 40\%.
By Equation~\eqref{eq:realized_utility}, this specification implies that a model with perfect predictions can achieve at most a realized utility of $\mathrm{SR}^2 / (2\gamma) = 4\%$.
As \citet{Bollerslev2018} describe, realized utility may be interpreted as the percentage of wealth the investor is willing to pay to have access to the model's forecasts; i.e., an investor would be willing to give up 4\% of their wealth to have access to the portfolio guided by a perfect model instead of investing entirely in the risk-free asset.
We assess the realized utility for each model on a per-stock basis and report the average utility across all stocks to provide a broad overview of the models' economic performance.
Moreover, we also conduct the realized utility analysis when transaction costs are considered.
In this setting, we follow \citet{Bollerslev2018} and use the rolling median of the bid-ask spread over the last nine months as a proxy for transaction costs.
\section{Results}\label{sec:results}
This section reports the main findings for the full sample of {\fullstocks} stocks.
Additional tables focusing solely on DJIA and Nasdaq-100 constituents are provided in \ref{sec:results_dow30} and \ref{sec:results_nasdaq100}, respectively.

\begin{table}[h!]
    \begin{center}
    \begin{tabular}{l cc c cc}
    \toprule
    & \multicolumn{2}{c}{\textbf{MSE}} & & \multicolumn{2}{c}{\textbf{QLIKE}} \\
    \cmidrule{2-3} \cmidrule{5-6}
    & \textbf{No VIX} & \textbf{VIX} & & \textbf{No VIX} & \textbf{VIX} \\
    \midrule    HAR (OLS) & 61.6\% & \textbf{88.5\%} & & 22.1\% & 38.6\%  \\
    HAR (WLS) & 37.9\% & 67.7\% & & 73.1\% & \textbf{85.5\%}  \\
    HAR (OLS, pooled) & 59.1\% & 68.4\% & & 11.6\% & 13.4\%  \\
    HAR (WLS, pooled) & 42.8\% & 65.2\% & & 56.5\% & 53.1\%  \\
    lasso & 58.3\% & 70.5\% & & 20.0\% & 70.9\%  \\
    Random Forest & 46.5\% & 50.5\% & & 25.4\% & 49.0\%  \\
    Gradient Boosted Trees & 35.1\% & 43.0\% & & 22.8\% & 48.4\%  \\
    Feedforward Neural Network & 11.1\% & 17.3\% & & 14.6\% & 36.4\%  \\
    \bottomrule
    \end{tabular}
    \caption{Percentage of assets (out of the full sample of {\fullstocks} stocks) for which the model is part of the best model class according to the MCS procedure with a 95\% confidence level.}
    \label{tbl:mcs_results}
    \end{center}
\end{table} 

Table~\ref{tbl:mcs_results} showcases the results of the MCS procedure using MSE and QLIKE as the loss functions, while Table~\ref{tbl:mse_all} and Table~\ref{tbl:qlike_all} provide detailed statistics for each model and loss function.
In particular, Table~\ref{tbl:mcs_results} details the percentage of assets for which each model is considered to belong to the best model class with a 95\% confidence level.
A notable observation is the consistent improvement in model performance when incorporating the VIX as an additional predictor, except for the QLIKE of the HAR model estimated via WLS in a pooled manner.
This performance gain is more pronounced in non-pooled models, aligning with the expectation that VIX, as a market-wide volatility measure, offers more value in models that do not already incorporate broad market trends through pooling.
Interestingly, this finding might also suggest that the benefits of pooling, as documented in \citet{Bollerslev2018,Zhang2023}, may diminish when the VIX is included in the model's information set.
This relationship between VIX inclusion and model performance underscores the importance of market-wide information in volatility forecasting. It corroborates previous findings about adding the VIX in the literature \citep{Audrino2020,Buncic2016,Zhang2020}.
The divergence in performance between the MSE and QLIKE metrics is particularly pronounced in the HAR models, which are estimated via OLS.
This discrepancy highlights the importance of selecting an appropriate estimation method for the HAR model, supporting the findings of \citet{Clements2021} and underscoring the potential pitfalls of using OLS to estimate the HAR model in situations where one aims to capture the asymmetries and tail risks that the QLIKE function is sensitive to.

\begin{table}[h!]
        \begin{center}
        % \resizebox{\columnwidth}{!}{
        \begin{tabular}{l ccccccc}
        \toprule
        & & & \multicolumn{5}{c}{\textbf{Quantiles}} \\
        \cmidrule(lr){4-8}
        \textbf{No VIX} & \textbf{Mean} & & \textbf{5\%} & \textbf{25\%} & \textbf{50\%} & \textbf{75\%} & \textbf{95\%} \\
        \midrule
    HAR (OLS) & 0.425 && 0.211 & 0.268 & 0.349 & 0.504 & 0.910\\
HAR (WLS) & 0.432 && 0.214 & 0.270 & 0.353 & 0.515 & 0.936\\
HAR (OLS, pooled) & 0.428 && 0.211 & 0.267 & 0.351 & 0.506 & 0.923\\
HAR (WLS, pooled) & 0.432 && 0.212 & 0.270 & 0.353 & 0.513 & 0.939\\
lasso & 0.432 && 0.213 & 0.273 & 0.356 & 0.516 & 0.921\\
Random Forest & 0.451 && 0.218 & 0.277 & 0.367 & 0.535 & 0.984\\
Gradient Boosted Trees & 0.463 && 0.219 & 0.282 & 0.371 & 0.548 & 1.022\\
Feedforward Neural Network & 0.559 && 0.234 & 0.306 & 0.441 & 0.674 & 1.280\\[0.5em]
    \textbf{VIX} \\
        \midrule
    HAR (OLS) & \textbf{0.419} && \textbf{0.204} & \textbf{0.261} & \textbf{0.341} & \textbf{0.501} & \textbf{0.907}\\
HAR (WLS) & 0.425 && 0.206 & 0.262 & 0.343 & 0.507 & 0.926\\
HAR (OLS, pooled) & 0.426 && 0.210 & 0.265 & 0.348 & 0.506 & 0.919\\
HAR (WLS, pooled) & 0.428 && 0.208 & 0.266 & 0.348 & 0.509 & 0.935\\
lasso & 0.435 && 0.206 & 0.267 & 0.352 & 0.527 & 0.928\\
Random Forest & 0.460 && 0.214 & 0.279 & 0.370 & 0.553 & 1.015\\
Gradient Boosted Trees & 0.473 && 0.216 & 0.279 & 0.377 & 0.567 & 1.057\\
Feedforward Neural Network & 0.568 && 0.228 & 0.306 & 0.435 & 0.684 & 1.280\\
    \bottomrule
        \end{tabular}%}
        \caption{Descriptive statistics for the MSEs of the different models without and with VIX. The table shows MSEs calculated during the test period from January 2022 to November 2023, averaged on a per-stock basis (out of the full sample of \fullstocks \ stocks). The lowest value in each column is highlighted in bold.}
        \label{tbl:mse_all}
        \end{center}
    \end{table}

A salient observation from our analysis is the relative underperformance of ML models compared to HAR models, especially when the latter are estimated using WLS.
Although the lasso model stands out with competitive performance, other ML models consistently lag behind the HAR model estimated via WLS by a substantial margin, regardless of the performance metric used.
This is further corroborated by the summary statistics in Tables~\ref{tbl:mse_all} and \ref{tbl:qlike_all}.
These findings contradict previous studies suggesting that ML models can outperform HAR models in volatility forecasting \citep{Zhang2023,Christensen2023,Qiu2021}.
As Figure~\ref{fig:har_heatmap} illustrates and as we argue throughout this work, the choice of fitting scheme for the HAR model is a crucial determinant of its forecasting accuracy and a major reason for the discrepancy in results across studies.
Moreover, the results we obtain for DJIA and Nasdaq-100 constituents in \ref{sec:results_dow30} and \ref{sec:results_nasdaq100} are consistent with the findings for the full sample, and the underperformance of ML models is even more pronounced in these subsets.

\begin{table}[h!]
        \begin{center}
        % \resizebox{\columnwidth}{!}{
        \begin{tabular}{l ccccccc}
        \toprule
        & & & \multicolumn{5}{c}{\textbf{Quantiles}} \\
        \cmidrule(lr){4-8}
        \textbf{No VIX} & \textbf{Mean} & & \textbf{5\%} & \textbf{25\%} & \textbf{50\%} & \textbf{75\%} & \textbf{95\%} \\
        \midrule
    HAR (OLS) & 0.333 && 0.122 & 0.170 & 0.250 & 0.400 & 0.806\\
HAR (WLS) & 0.314 && 0.120 & 0.167 & 0.240 & 0.379 & \textbf{0.741}\\
HAR (OLS, pooled) & 0.346 && 0.126 & 0.176 & 0.260 & 0.416 & 0.853\\
HAR (WLS, pooled) & 0.329 && 0.120 & 0.168 & 0.247 & 0.396 & 0.814\\
lasso & 0.341 && 0.128 & 0.184 & 0.266 & 0.412 & 0.810\\
Random Forest & 0.350 && 0.129 & 0.187 & 0.270 & 0.417 & 0.836\\
Gradient Boosted Trees & 0.350 && 0.133 & 0.190 & 0.274 & 0.419 & 0.824\\
Feedforward Neural Network & 0.477 && 0.140 & 0.205 & 0.310 & 0.522 & 1.088\\[0.5em]
    \textbf{VIX} \\
        \midrule
    HAR (OLS) & 0.332 && 0.117 & 0.170 & 0.248 & 0.399 & 0.814\\
HAR (WLS) & \textbf{0.313} && \textbf{0.115} & \textbf{0.164} & \textbf{0.239} & \textbf{0.377} & 0.743\\
HAR (OLS, pooled) & 0.350 && 0.127 & 0.180 & 0.264 & 0.421 & 0.853\\
HAR (WLS, pooled) & 0.332 && 0.120 & 0.169 & 0.251 & 0.401 & 0.809\\
lasso & 0.320 && 0.117 & 0.172 & 0.247 & 0.388 & 0.769\\
Random Forest & 0.344 && 0.122 & 0.182 & 0.262 & 0.405 & 0.825\\
Gradient Boosted Trees & 0.344 && 0.122 & 0.181 & 0.265 & 0.409 & 0.821\\
Feedforward Neural Network & 0.571 && 0.129 & 0.195 & 0.294 & 0.484 & 1.065\\
    \bottomrule
        \end{tabular}%}
        \caption{Descriptive statistics for the QLIKE of the different models without and with VIX. The table shows QLIKE calculated during the test period from January 2022 to November 2023, averaged on a per-stock basis (out of the full sample of \fullstocks \ stocks). The lowest value in each column is highlighted in bold.}
        \label{tbl:qlike_all}
        \end{center}
    \end{table}

Table~\ref{tbl:ru_all} and Table~\ref{tbl:rutc_all} illustrate the economic relevance of the models through the financial perspective of realized utility as defined in \citet{Bollerslev2018}, with and without considering transaction costs, respectively.
In both scenarios, the HAR model estimated via WLS markedly outperforms the other models, regardless of whether VIX is included.
As described in \citet{Bollerslev2018}, the values may be interpreted as the percentage of wealth that an investor would be willing to pay in order to access a portfolio that leverages the model's forecasting capabilities compared to a portfolio that uniquely allocates capital in the risk-free asset.
In turn, the difference in realized utilities between the various models can be interpreted as the wealth that an investor would be willing to pay to access the superior forecasting capabilities.
For instance, when transactions are costly and the VIX is excluded, an investor using the lasso model would be willing to pay 8.7 basis points of their wealth as a fee to access the forecasts of the HAR model estimated via WLS.
This consistent superiority in economic performance strengthens the findings obtained from the statistical losses and underscores the prevalence of the HAR model in volatility forecasting. 
Moreover, the results for realized utilities for DJIA and Nasdaq-100 constituents in \ref{sec:results_dow30} and \ref{sec:results_nasdaq100} confirm the robustness of these findings within subsets of more actively traded and economically significant stocks.

\begin{table}[h!]
    \begin{center}
    \resizebox{\columnwidth}{!}{
    \begin{tabular}{l ccccccc}
    \toprule
    & & & \multicolumn{5}{c}{\textbf{Quantiles}} \\
    \cmidrule(lr){4-8}
    \textbf{No VIX} & \textbf{Mean} & & \textbf{5\%} & \textbf{25\%} & \textbf{50\%} & \textbf{75\%} & \textbf{95\%} \\
    \midrule
HAR (OLS) & 3.152\% && 1.794\% & 2.992\% & 3.403\% & 3.615\% & 3.738\%\\
HAR (WLS) & \textbf{3.218}\% && 2.019\% & \textbf{3.067}\% & 3.435\% & \textbf{3.630}\% & 3.745\%\\
HAR (OLS, pooled) & 3.111\% && 1.685\% & 2.958\% & 3.373\% & 3.594\% & 3.724\%\\
HAR (WLS, pooled) & 3.168\% && 1.817\% & 3.027\% & 3.414\% & 3.621\% & 3.743\%\\
lasso & 3.131\% && 1.806\% & 2.968\% & 3.363\% & 3.573\% & 3.718\%\\
Random Forest & 3.114\% && 1.741\% & 2.958\% & 3.351\% & 3.565\% & 3.714\%\\
Gradient Boosted Trees & 3.119\% && 1.771\% & 2.964\% & 3.345\% & 3.561\% & 3.708\%\\
Feedforward Neural Network & 2.796\% && 1.058\% & 2.731\% & 3.259\% & 3.531\% & 3.693\%\\[0.5em]
    \textbf{VIX} \\
    \midrule
HAR (OLS) & 3.149\% && 1.794\% & 2.986\% & 3.408\% & 3.612\% & 3.747\%\\
HAR (WLS) & 3.217\% && \textbf{2.037}\% & 3.062\% & \textbf{3.439}\% & 3.629\% & \textbf{3.755}\%\\
HAR (OLS, pooled) & 3.095\% && 1.689\% & 2.933\% & 3.356\% & 3.581\% & 3.720\%\\
HAR (WLS, pooled) & 3.157\% && 1.821\% & 3.007\% & 3.400\% & 3.611\% & 3.741\%\\
lasso & 3.202\% && 1.928\% & 3.048\% & 3.413\% & 3.612\% & 3.752\%\\
Random Forest & 3.139\% && 1.757\% & 3.010\% & 3.376\% & 3.586\% & 3.736\%\\
Gradient Boosted Trees & 3.151\% && 1.786\% & 3.023\% & 3.369\% & 3.585\% & 3.740\%\\
Feedforward Neural Network & 2.367\% && 1.055\% & 2.810\% & 3.306\% & 3.560\% & 3.726\%\\
    \bottomrule
    \end{tabular}}
    \caption{Descriptive statistics for the realized utilities (without transaction costs) of the different models without and with VIX. The table shows realized utilities calculated during the test period from January 2022 to November 2023 for the full sample of \fullstocks \ stocks. The largest value in each column is highlighted in bold.}
    \label{tbl:ru_all}
    \end{center}
\end{table}
\begin{table}[h!]
    \begin{center}
    \resizebox{\columnwidth}{!}{
    \begin{tabular}{l ccccccc}
    \toprule
    & & & \multicolumn{5}{c}{\textbf{Quantiles}} \\
    \cmidrule(lr){4-8}
    \textbf{No VIX} & \textbf{Mean} & & \textbf{5\%} & \textbf{25\%} & \textbf{50\%} & \textbf{75\%} & \textbf{95\%} \\
    \midrule
HAR (OLS) & 2.273\% && -0.979\% & 2.326\% & 3.177\% & 3.482\% & 3.671\%\\
HAR (WLS) & \textbf{2.340}\% && \textbf{-0.733}\% & \textbf{2.404}\% & \textbf{3.205}\% & 3.499\% & 3.680\%\\
HAR (OLS, pooled) & 2.233\% && -0.934\% & 2.258\% & 3.141\% & 3.469\% & 3.654\%\\
HAR (WLS, pooled) & 2.290\% && -0.840\% & 2.336\% & 3.186\% & 3.496\% & 3.673\%\\
lasso & 2.253\% && -0.900\% & 2.286\% & 3.112\% & 3.452\% & 3.648\%\\
Random Forest & 2.236\% && -1.045\% & 2.281\% & 3.100\% & 3.443\% & 3.640\%\\
Gradient Boosted Trees & 2.241\% && -0.838\% & 2.253\% & 3.090\% & 3.435\% & 3.637\%\\
Feedforward Neural Network & 1.918\% && -1.518\% & 2.050\% & 2.982\% & 3.397\% & 3.623\%\\[0.5em]
    \textbf{VIX} \\
    \midrule
HAR (OLS) & 2.271\% && -1.035\% & 2.298\% & 3.163\% & 3.482\% & 3.679\%\\
HAR (WLS) & 2.338\% && -0.827\% & 2.376\% & 3.195\% & \textbf{3.501}\% & \textbf{3.688}\%\\
HAR (OLS, pooled) & 2.217\% && -0.990\% & 2.221\% & 3.120\% & 3.460\% & 3.643\%\\
HAR (WLS, pooled) & 2.278\% && -0.870\% & 2.306\% & 3.166\% & 3.488\% & 3.667\%\\
lasso & 2.323\% && -0.749\% & 2.373\% & 3.176\% & 3.483\% & 3.679\%\\
Random Forest & 2.261\% && -0.856\% & 2.301\% & 3.122\% & 3.460\% & 3.664\%\\
Gradient Boosted Trees & 2.273\% && -0.843\% & 2.296\% & 3.127\% & 3.458\% & 3.663\%\\
Feedforward Neural Network & 1.488\% && -1.352\% & 2.092\% & 3.033\% & 3.427\% & 3.648\%\\
    \bottomrule
    \end{tabular}}
    \caption{Descriptive statistics for the realized utilities (with transaction costs) of the different models without and with VIX. The table shows realized utilities calculated during the test period from January 2022 to November 2023 for the full sample of \fullstocks \ stocks. The largest value in each column is highlighted in bold.}
    \label{tbl:rutc_all}
    \end{center}
\end{table}

\begin{figure}[!htb]
    \centering
    \includegraphics[width=\textwidth]{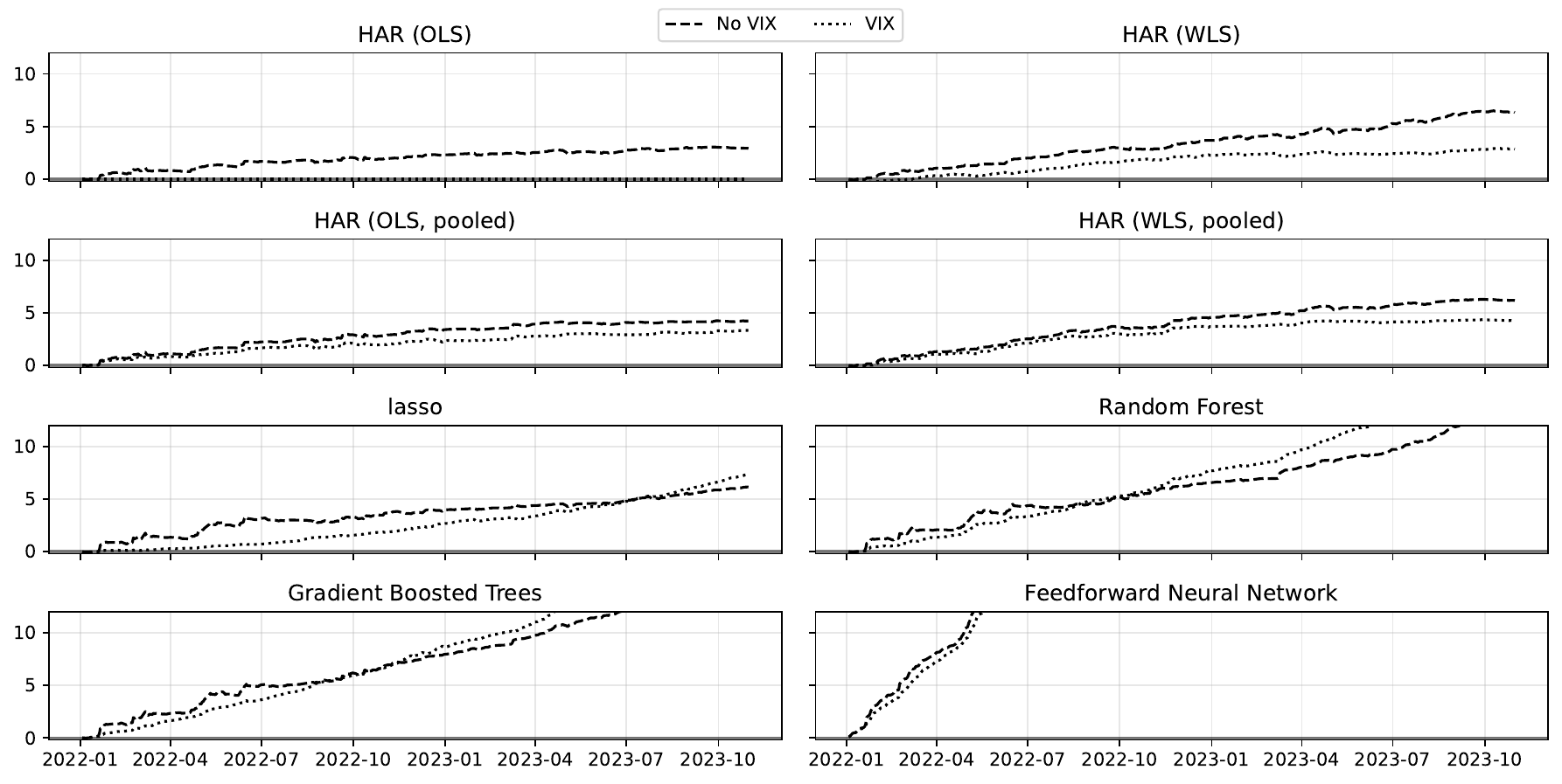}
    \caption{Cumulative squared error difference between the models and the HAR-VIX model estimated via OLS for all {\fullstocks} stocks. The dashed lines represent the models without the VIX, while the dotted lines represent the models when the VIX is included. For better visualization, the $y$-axis is constrained to the range $[0, 12]$.}
    \label{fig:cse_diff}
\end{figure}

Figure~\ref{fig:cse_diff} illustrates the cumulative squared error difference (CSED) for various models compared to the HAR-VIX model estimated via OLS, which provides the best performance in terms of average MSE across all assets.
The figure highlights that the difference in performance between the models is not driven by specific time spans but is consistent across the entire testing period from early 2022 to late 2023.
All HAR models benefit from the inclusion of the VIX in a consistent manner; i.e., throughout the entire time period, the CSED is lower when the VIX is included.
This observation does not hold for the ML models, which all exhibit a similar pattern where the CSED when the VIX is included is lower at the beginning of the testing period but increases over time and eventually surpasses the value of the CSED when the VIX is not included.
However, this pattern of performance deterioration over time is not observed in the restricted samples of DJIA and Nasdaq-100 constituents, as shown in in the appendices.
In these subsets, the CSED indicates that the inclusion of the VIX consistently improves the performance.
\section{Conclusion}\label{sec:conclusion}
Grounded on a substantial and unprecedented dataset of {\fullstocks} stocks from the U.S. equity market, our empirical analysis demonstrates that a carefully fitted HAR model remains a particularly robust benchmark for realized volatility forecasting.
Despite the growing popularity of ML techniques in financial econometrics and their usefulness in capturing complex nonlinear patterns, our results challenge the notion that these methods may be a universal solution for volatility prediction.
The additional model complexity and associated computational burden of traditional ML algorithms may negatively impact forecasting performance, especially when the information set is limited.

In contrast, the HAR model's linear structure makes it a lightweight, interpretable, and robust forecasting tool, particularly when the fitting scheme is meticulously selected.
Our results emphasize the necessity of carefully specifying the training window and re-estimation frequency, demonstrating that these critical specification choices significantly influence the HAR model's performance.

Even when tested across subsets of the DJIA and Nasdaq-100 constituents, our findings remain robust, reinforcing the HAR model's broad applicability and practical relevance in realized volatility forecasting.
\clearpage    
% % Bibliography.
\bibliography{bibliography}

\begin{thebibliography}{38}
\expandafter\ifx\csname natexlab\endcsname\relax\def\natexlab#1{#1}\fi
\providecommand{\url}[1]{\texttt{#1}}
\providecommand{\href}[2]{#2}
\providecommand{\path}[1]{#1}
\providecommand{\DOIprefix}{doi:}
\providecommand{\ArXivprefix}{arXiv:}
\providecommand{\URLprefix}{URL: }
\providecommand{\Pubmedprefix}{pmid:}
\providecommand{\doi}[1]{\href{http://dx.doi.org/#1}{\path{#1}}}
\providecommand{\Pubmed}[1]{\href{pmid:#1}{\path{#1}}}
\providecommand{\bibinfo}[2]{#2}
\ifx\xfnm\relax \def\xfnm[#1]{\unskip,\space#1}\fi
%Type = Article
\bibitem[{Andersen et~al.(2001)Andersen, Bollerslev, Diebold \&
  Labys}]{Andersen2001}
\bibinfo{author}{Andersen, T.~G.}, \bibinfo{author}{Bollerslev, T.},
  \bibinfo{author}{Diebold, F.~X.}, \& \bibinfo{author}{Labys, P.}
  (\bibinfo{year}{2001}).
\newblock \bibinfo{title}{{The distribution of exchange rate volatility}}.
\newblock {\it \bibinfo{journal}{Journal of the American Statistical
  Association}\/},  {\it \bibinfo{volume}{96}\/}, \bibinfo{pages}{42--55}.
%Type = Article
\bibitem[{Andersen et~al.(2003)Andersen, Bollerslev, Diebold \&
  Labys}]{Andersen2003}
\bibinfo{author}{Andersen, T.~G.}, \bibinfo{author}{Bollerslev, T.},
  \bibinfo{author}{Diebold, F.~X.}, \& \bibinfo{author}{Labys, P.}
  (\bibinfo{year}{2003}).
\newblock \bibinfo{title}{{Modeling and forecasting realized volatility}}.
\newblock {\it \bibinfo{journal}{Econometrica}\/},  {\it
  \bibinfo{volume}{71}\/}, \bibinfo{pages}{579--625}.
%Type = Article
\bibitem[{Audrino \& Knaus(2016)}]{Audrino2016}
\bibinfo{author}{Audrino, F.}, \& \bibinfo{author}{Knaus, S.~D.}
  (\bibinfo{year}{2016}).
\newblock \bibinfo{title}{{Lassoing the HAR Model: A Model Selection
  Perspective on Realized Volatility Dynamics}}.
\newblock {\it \bibinfo{journal}{Econometric Reviews}\/},  {\it
  \bibinfo{volume}{35}\/}, \bibinfo{pages}{1485--1521}.
%Type = Article
\bibitem[{Audrino et~al.(2020)Audrino, Sigrist \& Ballinari}]{Audrino2020}
\bibinfo{author}{Audrino, F.}, \bibinfo{author}{Sigrist, F.}, \&
  \bibinfo{author}{Ballinari, D.} (\bibinfo{year}{2020}).
\newblock \bibinfo{title}{{The impact of sentiment and attention measures on
  stock market volatility}}.
\newblock {\it \bibinfo{journal}{International Journal of Forecasting}\/},
  {\it \bibinfo{volume}{36}\/}, \bibinfo{pages}{334--357}.
%Type = Article
\bibitem[{Barndorff-Nielsen et~al.(2009)Barndorff-Nielsen, Hansen, Lunde \&
  Shephard}]{Barndorff-Nielsen2009}
\bibinfo{author}{Barndorff-Nielsen, O.~E.}, \bibinfo{author}{Hansen, P.~R.},
  \bibinfo{author}{Lunde, A.}, \& \bibinfo{author}{Shephard, N.}
  (\bibinfo{year}{2009}).
\newblock \bibinfo{title}{{Realized kernels in practice: trades and quotes}}.
\newblock {\it \bibinfo{journal}{The Econometrics Journal}\/},  {\it
  \bibinfo{volume}{12}\/}, \bibinfo{pages}{C1--C32}.
%Type = Article
\bibitem[{Barndorff-Nielsen \&
  Shephard(2002{\natexlab{a}})}]{Barndorff-Nielsen2002}
\bibinfo{author}{Barndorff-Nielsen, O.~E.}, \& \bibinfo{author}{Shephard, N.}
  (\bibinfo{year}{2002}{\natexlab{a}}).
\newblock \bibinfo{title}{{Econometric analysis of realized volatility and its
  use in estimating stochastic volatility models}}.
\newblock {\it \bibinfo{journal}{Journal of the Royal Statistical Society.
  Series B: Statistical Methodology}\/},  {\it \bibinfo{volume}{64}\/},
  \bibinfo{pages}{253--280}.
%Type = Article
\bibitem[{Barndorff-Nielsen \&
  Shephard(2002{\natexlab{b}})}]{Barndorff-Nielsen2002a}
\bibinfo{author}{Barndorff-Nielsen, O.~E.}, \& \bibinfo{author}{Shephard, N.}
  (\bibinfo{year}{2002}{\natexlab{b}}).
\newblock \bibinfo{title}{{Estimating quadratic variation using realized
  variance}}.
\newblock {\it \bibinfo{journal}{Journal of Applied Econometrics}\/},  {\it
  \bibinfo{volume}{17}\/}, \bibinfo{pages}{457--477}.
%Type = Book
\bibitem[{Bauwens et~al.(2012)Bauwens, Hafner \& Laurent}]{Bauwens2012}
\bibinfo{author}{Bauwens, L.}, \bibinfo{author}{Hafner, C.}, \&
  \bibinfo{author}{Laurent, S.} (\bibinfo{year}{2012}).
\newblock {\it \bibinfo{title}{{Volatility Models and Their Applications}}\/}
  volume~\bibinfo{volume}{15}.
\newblock \bibinfo{publisher}{John Wiley \& Sons, Inc.}
%Type = Article
\bibitem[{Bollerslev et~al.(2018)Bollerslev, Hood, Huss \&
  Pedersen}]{Bollerslev2018}
\bibinfo{author}{Bollerslev, T.}, \bibinfo{author}{Hood, B.},
  \bibinfo{author}{Huss, J.}, \& \bibinfo{author}{Pedersen, L.~H.}
  (\bibinfo{year}{2018}).
\newblock \bibinfo{title}{{Risk Everywhere: Modeling and managing volatility}}.
\newblock {\it \bibinfo{journal}{Review of Financial Studies}\/},  {\it
  \bibinfo{volume}{31}\/}, \bibinfo{pages}{2730--2773}.
%Type = Article
\bibitem[{Breiman(1996)}]{Breiman1996}
\bibinfo{author}{Breiman, L.} (\bibinfo{year}{1996}).
\newblock \bibinfo{title}{{Bagging Predictors}}.
\newblock {\it \bibinfo{journal}{Machine Learning}\/},  {\it
  \bibinfo{volume}{24}\/}, \bibinfo{pages}{123--140}.
%Type = Article
\bibitem[{Breiman(2001)}]{Breiman2001}
\bibinfo{author}{Breiman, L.} (\bibinfo{year}{2001}).
\newblock \bibinfo{title}{{Random Forests}}.
\newblock {\it \bibinfo{journal}{Machine Learning}\/},  {\it
  \bibinfo{volume}{45}\/}, \bibinfo{pages}{5--32}.
%Type = Article
\bibitem[{Bucci(2020)}]{Bucci2020}
\bibinfo{author}{Bucci, A.} (\bibinfo{year}{2020}).
\newblock \bibinfo{title}{{Realized Volatility Forecasting with Neural
  Networks}}.
\newblock {\it \bibinfo{journal}{Journal of Financial Econometrics}\/},  {\it
  \bibinfo{volume}{18}\/}, \bibinfo{pages}{502--531}.
%Type = Article
\bibitem[{Buncic \& Gisler(2016)}]{Buncic2016}
\bibinfo{author}{Buncic, D.}, \& \bibinfo{author}{Gisler, K.~I.}
  (\bibinfo{year}{2016}).
\newblock \bibinfo{title}{{Global equity market volatility spillovers: A
  broader role for the United States}}.
\newblock {\it \bibinfo{journal}{International Journal of Forecasting}\/},
  {\it \bibinfo{volume}{32}\/}, \bibinfo{pages}{1317--1339}.
%Type = Article
\bibitem[{Christensen et~al.(2023)Christensen, Siggaard \&
  Veliyev}]{Christensen2023}
\bibinfo{author}{Christensen, K.}, \bibinfo{author}{Siggaard, M.}, \&
  \bibinfo{author}{Veliyev, B.} (\bibinfo{year}{2023}).
\newblock \bibinfo{title}{{A Machine Learning Approach to Volatility
  Forecasting}}.
\newblock {\it \bibinfo{journal}{Journal of Financial Econometrics}\/},  {\it
  \bibinfo{volume}{21}\/}, \bibinfo{pages}{1680--1727}.
%Type = Article
\bibitem[{Cipollini et~al.(2021)Cipollini, Gallo \& Otranto}]{Cipollini2021}
\bibinfo{author}{Cipollini, F.}, \bibinfo{author}{Gallo, G.~M.}, \&
  \bibinfo{author}{Otranto, E.} (\bibinfo{year}{2021}).
\newblock \bibinfo{title}{{Realized volatility forecasting: Robustness to
  measurement errors}}.
\newblock {\it \bibinfo{journal}{International Journal of Forecasting}\/},
  {\it \bibinfo{volume}{37}\/}, \bibinfo{pages}{44--57}.
%Type = Article
\bibitem[{Clements \& Preve(2021)}]{Clements2021}
\bibinfo{author}{Clements, A.}, \& \bibinfo{author}{Preve, D.~P.}
  (\bibinfo{year}{2021}).
\newblock \bibinfo{title}{{A Practical Guide to harnessing the HAR volatility
  model}}.
\newblock {\it \bibinfo{journal}{Journal of Banking and Finance}\/},  {\it
  \bibinfo{volume}{133}\/}, \bibinfo{pages}{106285}.
%Type = Article
\bibitem[{Corsi(2009)}]{Corsi2009}
\bibinfo{author}{Corsi, F.} (\bibinfo{year}{2009}).
\newblock \bibinfo{title}{{A simple approximate long-memory model of realized
  volatility}}.
\newblock {\it \bibinfo{journal}{Journal of Financial Econometrics}\/},  {\it
  \bibinfo{volume}{7}\/}, \bibinfo{pages}{174--196}.
%Type = Incollection
\bibitem[{Corsi et~al.(2012)Corsi, Audrino \& Ren{\`{o}}}]{Corsi2012}
\bibinfo{author}{Corsi, F.}, \bibinfo{author}{Audrino, F.}, \&
  \bibinfo{author}{Ren{\`{o}}, R.} (\bibinfo{year}{2012}).
\newblock \bibinfo{title}{{HAR Modeling for Realized Volatility Forecasting}}.
\newblock In {\it \bibinfo{booktitle}{Handbook of Volatility Models and Their
  Applications}\/} (pp. \bibinfo{pages}{363--382}).
\newblock \bibinfo{publisher}{John Wiley \& Sons, Ltd}.
%Type = Article
\bibitem[{Friedman(2001)}]{Friedman2001}
\bibinfo{author}{Friedman, J.~H.} (\bibinfo{year}{2001}).
\newblock \bibinfo{title}{{Greedy function approximation: A gradient boosting
  machine}}.
\newblock {\it \bibinfo{journal}{Annals of Statistics}\/},  {\it
  \bibinfo{volume}{29}\/}, \bibinfo{pages}{1189--1232}.
%Type = Book
\bibitem[{Goodfellow et~al.(2016)Goodfellow, Bengio \&
  Courville}]{Goodfellow2016}
\bibinfo{author}{Goodfellow, I.}, \bibinfo{author}{Bengio, Y.}, \&
  \bibinfo{author}{Courville, A.} (\bibinfo{year}{2016}).
\newblock {\it \bibinfo{title}{{Deep Learning}}\/}.
\newblock \bibinfo{address}{Singapore}: \bibinfo{publisher}{MIT Press}.
%Type = Article
\bibitem[{Gunnarsson et~al.(2024)Gunnarsson, Isern, Kaloudis, Risstad, Vigdel
  \& Westgaard}]{Gunnarsson2024}
\bibinfo{author}{Gunnarsson, E.~S.}, \bibinfo{author}{Isern, H.~R.},
  \bibinfo{author}{Kaloudis, A.}, \bibinfo{author}{Risstad, M.},
  \bibinfo{author}{Vigdel, B.}, \& \bibinfo{author}{Westgaard, S.}
  (\bibinfo{year}{2024}).
\newblock \bibinfo{title}{{Prediction of realized volatility and implied
  volatility indices using AI and machine learning: A review}}.
\newblock {\it \bibinfo{journal}{International Review of Financial
  Analysis}\/},  {\it \bibinfo{volume}{93}\/}, \bibinfo{pages}{103221}.
%Type = Article
\bibitem[{Hansen et~al.(2011)Hansen, Lunde \& Nason}]{Hansen2011}
\bibinfo{author}{Hansen, B. P.~R.}, \bibinfo{author}{Lunde, A.}, \&
  \bibinfo{author}{Nason, J.~M.} (\bibinfo{year}{2011}).
\newblock \bibinfo{title}{{The Model Confidence Set}}.
\newblock {\it \bibinfo{journal}{Econometrica}\/},  {\it
  \bibinfo{volume}{79}\/}, \bibinfo{pages}{453--497}.
%Type = Book
\bibitem[{Hastie et~al.(2009)Hastie, Tibshirani \& Friedman}]{Hastie2009}
\bibinfo{author}{Hastie, T.}, \bibinfo{author}{Tibshirani, R.}, \&
  \bibinfo{author}{Friedman, J.} (\bibinfo{year}{2009}).
\newblock {\it \bibinfo{title}{{The Elements of Statistical Learning}}\/}.
\newblock Springer Series in Statistics (\bibinfo{edition}{2nd} ed.).
\newblock \bibinfo{address}{New York, NY}: \bibinfo{publisher}{Springer New
  York}.
%Type = Book
\bibitem[{Hastie et~al.(2015)Hastie, Tibshirani \& Wainwright}]{Hastie2015}
\bibinfo{author}{Hastie, T.}, \bibinfo{author}{Tibshirani, R.}, \&
  \bibinfo{author}{Wainwright, M.} (\bibinfo{year}{2015}).
\newblock {\it \bibinfo{title}{{Statistical Learning with Sparsity}}\/} volume
  \bibinfo{volume}{143}.
\newblock \bibinfo{publisher}{Chapman and Hall/CRC}.
%Type = Article
\bibitem[{Hoerl \& Kennard(1970)}]{Hoerl1970}
\bibinfo{author}{Hoerl, A.~E.}, \& \bibinfo{author}{Kennard, R.~W.}
  (\bibinfo{year}{1970}).
\newblock \bibinfo{title}{{Ridge Regression: Biased Estimation for
  Nonorthogonal Problems}}.
\newblock {\it \bibinfo{journal}{Technometrics}\/},  {\it
  \bibinfo{volume}{12}\/}, \bibinfo{pages}{55--67}.
%Type = Article
\bibitem[{Hornik(1991)}]{Hornik1991}
\bibinfo{author}{Hornik, K.} (\bibinfo{year}{1991}).
\newblock \bibinfo{title}{{Approximation capabilities of multilayer feedforward
  networks}}.
\newblock {\it \bibinfo{journal}{Neural Networks}\/},  {\it
  \bibinfo{volume}{4}\/}, \bibinfo{pages}{251--257}.
%Type = Article
\bibitem[{Liu et~al.(2018)Liu, Pantelous \& von Mettenheim}]{Liu2018}
\bibinfo{author}{Liu, F.}, \bibinfo{author}{Pantelous, A.~A.}, \&
  \bibinfo{author}{von Mettenheim, H.~J.} (\bibinfo{year}{2018}).
\newblock \bibinfo{title}{{Forecasting and trading high frequency volatility on
  large indices}}.
\newblock {\it \bibinfo{journal}{Quantitative Finance}\/},  {\it
  \bibinfo{volume}{18}\/}, \bibinfo{pages}{737--748}.
%Type = Article
\bibitem[{Liu et~al.(2015)Liu, Patton \& Sheppard}]{Liu2015}
\bibinfo{author}{Liu, L.~Y.}, \bibinfo{author}{Patton, A.~J.}, \&
  \bibinfo{author}{Sheppard, K.} (\bibinfo{year}{2015}).
\newblock \bibinfo{title}{{Does anything beat 5-minute RV? A comparison of
  realized measures across multiple asset classes}}.
\newblock {\it \bibinfo{journal}{Journal of Econometrics}\/},  {\it
  \bibinfo{volume}{187}\/}, \bibinfo{pages}{293--311}.
%Type = Article
\bibitem[{Luong \& Dokuchaev(2018)}]{Luong2018}
\bibinfo{author}{Luong, C.}, \& \bibinfo{author}{Dokuchaev, N.}
  (\bibinfo{year}{2018}).
\newblock \bibinfo{title}{{Forecasting of Realised Volatility with the Random
  Forests Algorithm}}.
\newblock {\it \bibinfo{journal}{Journal of Risk and Financial Management}\/},
  {\it \bibinfo{volume}{11}\/}, \bibinfo{pages}{61}.
%Type = Article
\bibitem[{Patton(2011)}]{Patton2011}
\bibinfo{author}{Patton, A.~J.} (\bibinfo{year}{2011}).
\newblock \bibinfo{title}{{Volatility forecast comparison using imperfect
  volatility proxies}}.
\newblock {\it \bibinfo{journal}{Journal of Econometrics}\/},  {\it
  \bibinfo{volume}{160}\/}, \bibinfo{pages}{246--256}.
%Type = Article
\bibitem[{Patton \& Sheppard(2015)}]{Patton2015}
\bibinfo{author}{Patton, A.~J.}, \& \bibinfo{author}{Sheppard, K.}
  (\bibinfo{year}{2015}).
\newblock \bibinfo{title}{{Good volatility, bad volatility: Signed jumps and
  the persistence of volatility}}.
\newblock {\it \bibinfo{journal}{Review of Economics and Statistics}\/},  {\it
  \bibinfo{volume}{97}\/}, \bibinfo{pages}{683--697}.
%Type = Article
\bibitem[{Qiu(2021)}]{Qiu2021}
\bibinfo{author}{Qiu, Y.} (\bibinfo{year}{2021}).
\newblock \bibinfo{title}{{Complete subset least squares support vector
  regression}}.
\newblock {\it \bibinfo{journal}{Economics Letters}\/},  {\it
  \bibinfo{volume}{200}\/}, \bibinfo{pages}{109737}.
%Type = Book
\bibitem[{Takahashi et~al.(2023)Takahashi, Omori \& Watanabe}]{Takahashi2023}
\bibinfo{author}{Takahashi, M.}, \bibinfo{author}{Omori, Y.}, \&
  \bibinfo{author}{Watanabe, T.} (\bibinfo{year}{2023}).
\newblock {\it \bibinfo{title}{{Stochastic Volatility and Realized Stochastic
  Volatility Models}}\/}.
\newblock SpringerBriefs in Statistics.
\newblock \bibinfo{publisher}{Springer Nature}.
%Type = Article
\bibitem[{Taylor(2017)}]{Taylor2017}
\bibinfo{author}{Taylor, N.} (\bibinfo{year}{2017}).
\newblock \bibinfo{title}{{Realised variance forecasting under Box-Cox
  transformations}}.
\newblock {\it \bibinfo{journal}{International Journal of Forecasting}\/},
  {\it \bibinfo{volume}{33}\/}, \bibinfo{pages}{770--785}.
%Type = Article
\bibitem[{Tibshirani(1996)}]{Tibshirani1996}
\bibinfo{author}{Tibshirani, R.} (\bibinfo{year}{1996}).
\newblock \bibinfo{title}{{Regression Shrinkage and Selection Via the Lasso}}.
\newblock {\it \bibinfo{journal}{Journal of the Royal Statistical Society:
  Series B (Methodological)}\/},  {\it \bibinfo{volume}{58}\/},
  \bibinfo{pages}{267--288}.
%Type = Article
\bibitem[{Zhang et~al.(2023)Zhang, Zhang, Cucuringu \& Qian}]{Zhang2023}
\bibinfo{author}{Zhang, C.}, \bibinfo{author}{Zhang, Y.},
  \bibinfo{author}{Cucuringu, M.}, \& \bibinfo{author}{Qian, Z.}
  (\bibinfo{year}{2023}).
\newblock \bibinfo{title}{{Volatility Forecasting with Machine Learning and
  Intraday Commonality}}.
\newblock {\it \bibinfo{journal}{Journal of Financial Econometrics}\/},  (pp.
  \bibinfo{pages}{1--39}).
%Type = Article
\bibitem[{Zhang et~al.(2020)Zhang, Ma \& Liao}]{Zhang2020}
\bibinfo{author}{Zhang, Y.}, \bibinfo{author}{Ma, F.}, \&
  \bibinfo{author}{Liao, Y.} (\bibinfo{year}{2020}).
\newblock \bibinfo{title}{{Forecasting global equity market volatilities}}.
\newblock {\it \bibinfo{journal}{International Journal of Forecasting}\/},
  {\it \bibinfo{volume}{36}\/}, \bibinfo{pages}{1454--1475}.
%Type = Article
\bibitem[{Zhu et~al.(2023)Zhu, Bai, He \& Liu}]{Zhu2023}
\bibinfo{author}{Zhu, H.}, \bibinfo{author}{Bai, L.}, \bibinfo{author}{He, L.},
  \& \bibinfo{author}{Liu, Z.} (\bibinfo{year}{2023}).
\newblock \bibinfo{title}{{Forecasting realized volatility with machine
  learning: Panel data perspective}}.
\newblock {\it \bibinfo{journal}{Journal of Empirical Finance}\/},  {\it
  \bibinfo{volume}{73}\/}, \bibinfo{pages}{251--271}.

\end{thebibliography}
\clearpage
\appendix
\setcounter{table}{0}
\renewcommand{\thetable}{A.\arabic{table}}
\section{Expanding Window Figure}\label{sec:expanding_window}

\begin{figure}[!htb]
    \centering
    \includegraphics[width=\textwidth]{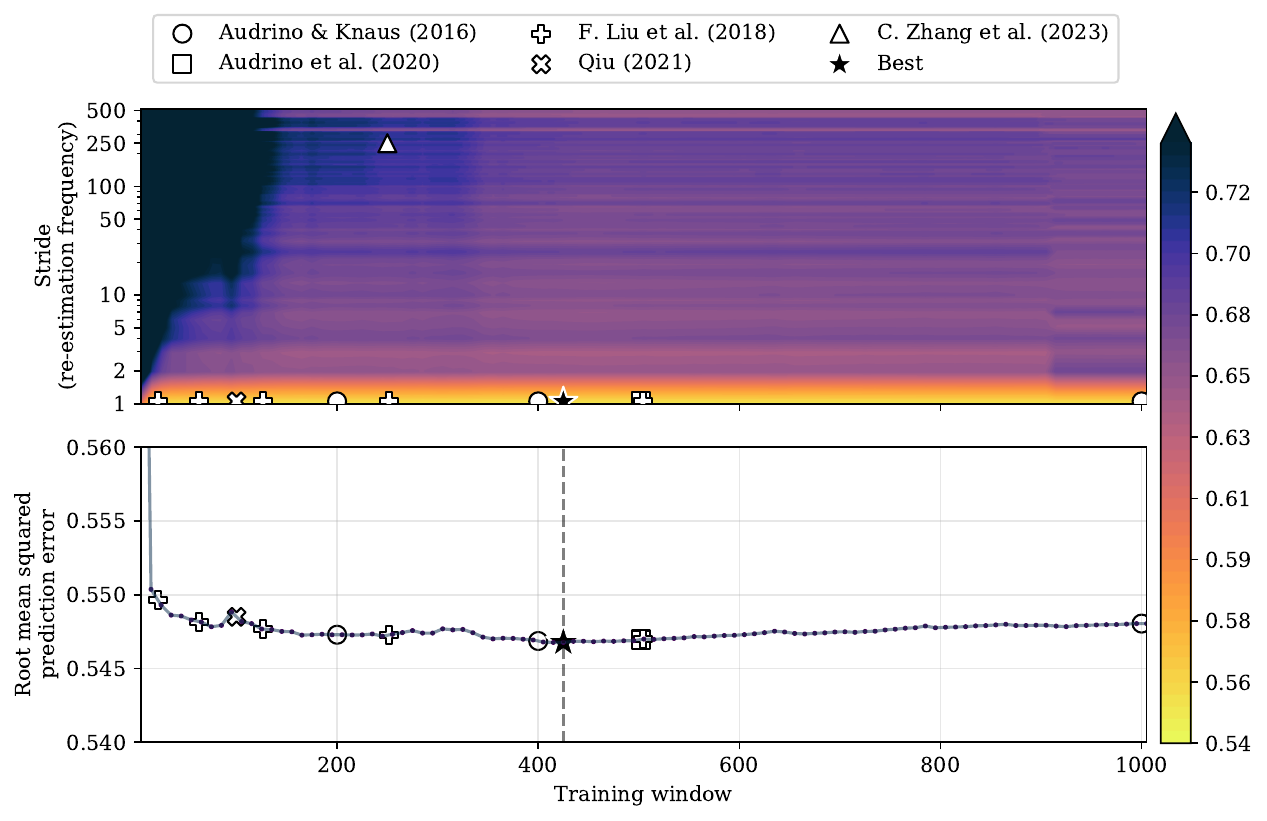}
    \caption{Top: Heatmap displaying the average root mean squared error (RMSE) of a HAR model with different fitting schemes for {\fullstocks} assets from January 2016 to December 2021. The $x$-axis shows the training window size, while the $y$-axis denotes the stride or re-estimation frequency. Color intensity varies with RMSE levels; darker colors indicate higher prediction errors. The fitting schemes used in previous studies are marked with distinct symbols for comparison. \\
    Bottom: Focused analysis on the scenario where the stride is $1$ (which emerges as the best-performing stride in the top plot). This plot provides a detailed look at how the size of the training window affects the RMSE, illustrating how larger training windows generally lead to lower prediction errors. \\
    In contrast to Figure~\ref{fig:har_heatmap}, this figure showcases the performance of the HAR model with an expanding window rather than a rolling window.}
    \label{fig:har_heatmap_expanding}
\end{figure}
\setcounter{table}{0}
\clearpage
\renewcommand{\thetable}{B.\arabic{table}}
\section{Results for the DJIA Constituents}\label{sec:results_dow30}

\begin{table}[h!]
    \begin{center}
    \begin{tabular}{l cc c cc}
    \toprule
    & \multicolumn{2}{c}{\textbf{MSE}} & & \multicolumn{2}{c}{\textbf{QLIKE}} \\
    \cmidrule{2-3} \cmidrule{5-6}
    & \textbf{No VIX} & \textbf{VIX} & & \textbf{No VIX} & \textbf{VIX} \\
    \midrule    HAR (OLS) & 44.4\% & 74.1\% & & 14.8\% & 18.5\%  \\
    HAR (WLS) & 29.6\% & \textbf{100.0\%} & & 74.1\% & \textbf{81.5\%}  \\
    HAR (OLS, pooled) & 22.2\% & 48.1\% & & 7.4\% & 14.8\%  \\
    HAR (WLS, pooled) & 7.4\% & 40.7\% & & 37.0\% & 51.9\%  \\
    lasso & 18.5\% & 66.7\% & & 7.4\% & 44.4\%  \\
    Random Forest & 18.5\% & 55.6\% & & 0.0\% & 14.8\%  \\
    Gradient Boosted Trees & 11.1\% & 51.9\% & & 3.7\% & 33.3\%  \\
    Feedforward Neural Network & 0.0\% & 11.1\% & & 3.7\% & 18.5\%  \\
    \bottomrule
    \end{tabular}
    \caption{Percentage of assets (within the DJIA stocks) for which the model is part of the best model class according to the MCS procedure with a 95\% confidence level.}
    \label{tbl:mcs_results_dow30}
    \end{center}
\end{table}
\begin{table}[h!]
        \begin{center}
        % \resizebox{\columnwidth}{!}{
        \begin{tabular}{l ccccccc}
        \toprule
        & & & \multicolumn{5}{c}{\textbf{Quantiles}} \\
        \cmidrule(lr){4-8}
        \textbf{No VIX} & \textbf{Mean} & & \textbf{5\%} & \textbf{25\%} & \textbf{50\%} & \textbf{75\%} & \textbf{95\%} \\
        \midrule
    HAR (OLS) & 0.248 && 0.209 & 0.232 & 0.243 & 0.262 & 0.293\\
HAR (WLS) & 0.250 && 0.210 & 0.235 & 0.246 & 0.263 & 0.294\\
HAR (OLS, pooled) & 0.252 && 0.210 & 0.236 & 0.249 & 0.267 & 0.293\\
HAR (WLS, pooled) & 0.254 && 0.212 & 0.238 & 0.251 & 0.269 & 0.295\\
lasso & 0.260 && 0.223 & 0.243 & 0.258 & 0.273 & 0.297\\
Random Forest & 0.254 && 0.215 & 0.241 & 0.256 & 0.266 & 0.297\\
Gradient Boosted Trees & 0.263 && 0.220 & 0.241 & 0.260 & 0.279 & 0.308\\
Feedforward Neural Network & 0.332 && 0.235 & 0.258 & 0.274 & 0.297 & 0.719\\[0.5em]
    \textbf{VIX} \\
        \midrule
    HAR (OLS) & 0.237 && 0.196 & \textbf{0.223} & 0.234 & 0.245 & 0.288\\
HAR (WLS) & \textbf{0.236} && \textbf{0.195} & 0.223 & \textbf{0.233} & \textbf{0.245} & \textbf{0.287}\\
HAR (OLS, pooled) & 0.247 && 0.209 & 0.230 & 0.241 & 0.264 & 0.290\\
HAR (WLS, pooled) & 0.247 && 0.211 & 0.230 & 0.246 & 0.264 & 0.288\\
lasso & 0.242 && 0.200 & 0.226 & 0.239 & 0.249 & 0.290\\
Random Forest & 0.248 && 0.210 & 0.233 & 0.241 & 0.265 & 0.294\\
Gradient Boosted Trees & 0.251 && 0.208 & 0.236 & 0.249 & 0.258 & 0.304\\
Feedforward Neural Network & 0.316 && 0.231 & 0.243 & 0.265 & 0.324 & 0.629\\
    \bottomrule
        \end{tabular}%}
        \caption{Descriptive statistics for the MSEs of the different models without and with VIX. The table shows MSEs calculated during the test period from January 2022 to November 2023, averaged on a per-stock basis (within the DJIA constituents). The lowest value in each column is highlighted in bold.}
        \label{tbl:mse_dow30}
        \end{center}
    \end{table}
\begin{table}[h!]
        \begin{center}
        % \resizebox{\columnwidth}{!}{
        \begin{tabular}{l ccccccc}
        \toprule
        & & & \multicolumn{5}{c}{\textbf{Quantiles}} \\
        \cmidrule(lr){4-8}
        \textbf{No VIX} & \textbf{Mean} & & \textbf{5\%} & \textbf{25\%} & \textbf{50\%} & \textbf{75\%} & \textbf{95\%} \\
        \midrule
    HAR (OLS) & 0.161 && 0.124 & 0.133 & 0.153 & 0.175 & 0.226\\
HAR (WLS) & \textbf{0.157} && 0.121 & \textbf{0.131} & 0.150 & \textbf{0.169} & \textbf{0.219}\\
HAR (OLS, pooled) & 0.164 && 0.122 & 0.140 & 0.158 & 0.176 & 0.227\\
HAR (WLS, pooled) & 0.160 && 0.119 & 0.136 & 0.154 & 0.172 & 0.221\\
lasso & 0.180 && 0.139 & 0.157 & 0.174 & 0.196 & 0.263\\
Random Forest & 0.176 && 0.133 & 0.149 & 0.167 & 0.197 & 0.238\\
Gradient Boosted Trees & 0.183 && 0.137 & 0.156 & 0.174 & 0.208 & 0.252\\
Feedforward Neural Network & 0.237 && 0.144 & 0.156 & 0.177 & 0.209 & 0.668\\[0.5em]
    \textbf{VIX} \\
        \midrule
    HAR (OLS) & 0.162 && 0.116 & 0.137 & 0.152 & 0.181 & 0.249\\
HAR (WLS) & 0.157 && \textbf{0.113} & 0.131 & \textbf{0.148} & 0.174 & 0.239\\
HAR (OLS, pooled) & 0.166 && 0.120 & 0.144 & 0.158 & 0.186 & 0.236\\
HAR (WLS, pooled) & 0.161 && 0.118 & 0.139 & 0.153 & 0.180 & 0.228\\
lasso & 0.164 && 0.121 & 0.142 & 0.155 & 0.176 & 0.256\\
Random Forest & 0.168 && 0.125 & 0.139 & 0.160 & 0.186 & 0.238\\
Gradient Boosted Trees & 0.169 && 0.121 & 0.152 & 0.160 & 0.181 & 0.255\\
Feedforward Neural Network & 0.220 && 0.134 & 0.156 & 0.178 & 0.197 & 0.493\\
    \bottomrule
        \end{tabular}%}
        \caption{Descriptive statistics for the QLIKE of the different models without and with VIX. The table shows QLIKE calculated during the test period from January 2022 to November 2023, averaged on a per-stock basis (within the DJIA constituents). The lowest value in each column is highlighted in bold.}
        \label{tbl:qlike_dow30}
        \end{center}
    \end{table}
\begin{table}[h!]
    \begin{center}
    \resizebox{\columnwidth}{!}{
    \begin{tabular}{l ccccccc}
    \toprule
    & & & \multicolumn{5}{c}{\textbf{Quantiles}} \\
    \cmidrule(lr){4-8}
    \textbf{No VIX} & \textbf{Mean} & & \textbf{5\%} & \textbf{25\%} & \textbf{50\%} & \textbf{75\%} & \textbf{95\%} \\
    \midrule
HAR (OLS) & 3.629\% && 3.427\% & 3.593\% & 3.663\% & 3.710\% & 3.733\%\\
HAR (WLS) & \textbf{3.642}\% && 3.449\% & \textbf{3.608}\% & \textbf{3.675}\% & \textbf{3.720}\% & 3.742\%\\
HAR (OLS, pooled) & 3.621\% && 3.435\% & 3.590\% & 3.642\% & 3.691\% & 3.733\%\\
HAR (WLS, pooled) & 3.635\% && \textbf{3.455}\% & 3.605\% & 3.655\% & 3.704\% & 3.743\%\\
lasso & 3.574\% && 3.297\% & 3.538\% & 3.601\% & 3.645\% & 3.699\%\\
Random Forest & 3.585\% && 3.389\% & 3.528\% & 3.620\% & 3.668\% & 3.707\%\\
Gradient Boosted Trees & 3.566\% && 3.362\% & 3.500\% & 3.588\% & 3.647\% & 3.697\%\\
Feedforward Neural Network & 3.434\% && 2.268\% & 3.510\% & 3.601\% & 3.652\% & 3.686\%\\[0.5em]
    \textbf{VIX} \\
    \midrule
HAR (OLS) & 3.617\% && 3.349\% & 3.579\% & 3.652\% & 3.695\% & 3.743\%\\
HAR (WLS) & 3.633\% && 3.377\% & 3.600\% & 3.666\% & 3.712\% & \textbf{3.752}\%\\
HAR (OLS, pooled) & 3.610\% && 3.406\% & 3.563\% & 3.636\% & 3.677\% & 3.737\%\\
HAR (WLS, pooled) & 3.627\% && 3.431\% & 3.581\% & 3.653\% & 3.693\% & 3.747\%\\
lasso & 3.616\% && 3.334\% & 3.588\% & 3.642\% & 3.684\% & 3.741\%\\
Random Forest & 3.606\% && 3.403\% & 3.563\% & 3.631\% & 3.696\% & 3.723\%\\
Gradient Boosted Trees & 3.604\% && 3.360\% & 3.579\% & 3.632\% & 3.657\% & 3.741\%\\
Feedforward Neural Network & 3.480\% && 2.783\% & 3.546\% & 3.596\% & 3.652\% & 3.710\%\\
    \bottomrule
    \end{tabular}}
    \caption{Descriptive statistics for the realized utilities (without transaction costs) of the different models without and with VIX. The table shows realized utilities calculated during the test period from January 2022 to November 2023 for the DJIA constituents. The largest value in each column is highlighted in bold.}
    \label{tbl:ru_dow30}
    \end{center}
\end{table}
\begin{table}[h!]
    \begin{center}
    \resizebox{\columnwidth}{!}{
    \begin{tabular}{l ccccccc}
    \toprule
    & & & \multicolumn{5}{c}{\textbf{Quantiles}} \\
    \cmidrule(lr){4-8}
    \textbf{No VIX} & \textbf{Mean} & & \textbf{5\%} & \textbf{25\%} & \textbf{50\%} & \textbf{75\%} & \textbf{95\%} \\
    \midrule
HAR (OLS) & 3.432\% && 3.011\% & 3.320\% & 3.510\% & 3.575\% & 3.636\%\\
HAR (WLS) & \textbf{3.445}\% && 3.026\% & 3.333\% & 3.524\% & 3.587\% & \textbf{3.646}\%\\
HAR (OLS, pooled) & 3.424\% && 3.012\% & 3.307\% & 3.526\% & 3.552\% & 3.625\%\\
HAR (WLS, pooled) & 3.438\% && 3.030\% & 3.321\% & \textbf{3.538}\% & 3.567\% & 3.635\%\\
lasso & 3.377\% && 2.993\% & 3.276\% & 3.463\% & 3.530\% & 3.591\%\\
Random Forest & 3.388\% && 2.948\% & 3.266\% & 3.459\% & 3.545\% & 3.605\%\\
Gradient Boosted Trees & 3.369\% && 2.962\% & 3.275\% & 3.431\% & 3.516\% & 3.600\%\\
Feedforward Neural Network & 3.237\% && 2.062\% & 3.176\% & 3.455\% & 3.528\% & 3.585\%\\[0.5em]
    \textbf{VIX} \\
    \midrule
HAR (OLS) & 3.420\% && 2.961\% & 3.326\% & 3.504\% & 3.581\% & 3.634\%\\
HAR (WLS) & 3.436\% && 2.983\% & \textbf{3.338}\% & 3.532\% & \textbf{3.593}\% & 3.643\%\\
HAR (OLS, pooled) & 3.414\% && 2.999\% & 3.311\% & 3.493\% & 3.537\% & 3.632\%\\
HAR (WLS, pooled) & 3.430\% && 3.019\% & 3.325\% & 3.518\% & 3.551\% & 3.641\%\\
lasso & 3.419\% && \textbf{3.032}\% & 3.326\% & 3.487\% & 3.565\% & 3.642\%\\
Random Forest & 3.409\% && 2.993\% & 3.303\% & 3.468\% & 3.565\% & 3.614\%\\
Gradient Boosted Trees & 3.407\% && 3.008\% & 3.318\% & 3.487\% & 3.534\% & 3.636\%\\
Feedforward Neural Network & 3.283\% && 2.651\% & 3.099\% & 3.410\% & 3.532\% & 3.596\%\\
    \bottomrule
    \end{tabular}}
    \caption{Descriptive statistics for the realized utilities (with transaction costs) of the different models without and with VIX. The table shows realized utilities calculated during the test period from January 2022 to November 2023 for the DJIA constituents. The largest value in each column is highlighted in bold.}
    \label{tbl:rutc_dow30}
    \end{center}
\end{table}

\begin{figure}[!htb]
    \centering
    \includegraphics[width=\textwidth]{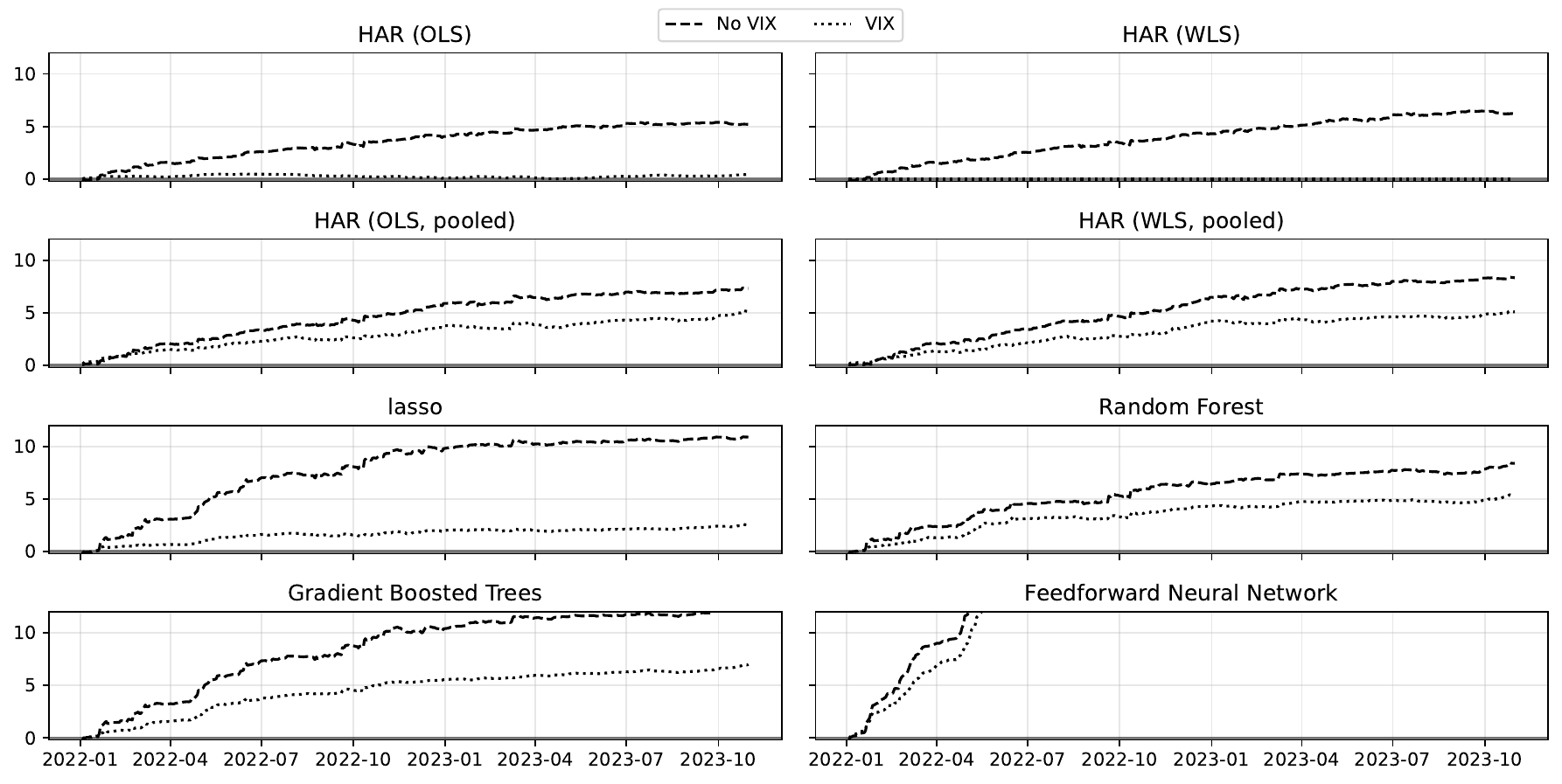}
    \caption{Cumulative squared error difference between the models and the HAR-VIX model estimated via WLS for the DJIA constituents. The dashed lines represent the models without the VIX, while the dotted lines represent the models when the VIX is included. For better visualization, the $y$-axis is constrained to the range $[0, 12]$.}
    \label{fig:cse_diff_dow30}
\end{figure}
\setcounter{table}{0}
\clearpage
\renewcommand{\thetable}{C.\arabic{table}}
\section{Results for the Nasdaq-100 Constituents}\label{sec:results_nasdaq100}

\begin{table}[h!]
    \begin{center}
    \begin{tabular}{l cc c cc}
    \toprule
    & \multicolumn{2}{c}{\textbf{MSE}} & & \multicolumn{2}{c}{\textbf{QLIKE}} \\
    \cmidrule{2-3} \cmidrule{5-6}
    & \textbf{No VIX} & \textbf{VIX} & & \textbf{No VIX} & \textbf{VIX} \\
    \midrule    HAR (OLS) & 55.7\% & 77.1\% & & 7.1\% & 12.9\%  \\
    HAR (WLS) & 50.0\% & \textbf{92.9\%} & & \textbf{87.1\%} & 67.1\%  \\
    HAR (OLS, pooled) & 38.6\% & 62.9\% & & 4.3\% & 7.1\%  \\
    HAR (WLS, pooled) & 34.3\% & 60.0\% & & 67.1\% & 65.7\%  \\
    lasso & 35.7\% & 70.0\% & & 7.1\% & 42.9\%  \\
    Random Forest & 31.4\% & 48.6\% & & 5.7\% & 24.3\%  \\
    Gradient Boosted Trees & 22.9\% & 48.6\% & & 5.7\% & 25.7\%  \\
    Feedforward Neural Network & 8.6\% & 12.9\% & & 5.7\% & 12.9\%  \\
    \bottomrule
    \end{tabular}
    \caption{Percentage of assets (within the Nasdaq-100 stocks) for which the model is part of the best model class according to the MCS procedure with a 95\% confidence level.}
    \label{tbl:mcs_results_nasdaq100}
    \end{center}
\end{table}
\begin{table}[h!]
        \begin{center}
        % \resizebox{\columnwidth}{!}{
        \begin{tabular}{l ccccccc}
        \toprule
        & & & \multicolumn{5}{c}{\textbf{Quantiles}} \\
        \cmidrule(lr){4-8}
        \textbf{No VIX} & \textbf{Mean} & & \textbf{5\%} & \textbf{25\%} & \textbf{50\%} & \textbf{75\%} & \textbf{95\%} \\
        \midrule
    HAR (OLS) & 0.281 && 0.206 & 0.233 & 0.262 & 0.310 & 0.430\\
HAR (WLS) & 0.283 && 0.209 & 0.234 & 0.263 & 0.312 & 0.438\\
HAR (OLS, pooled) & 0.284 && 0.209 & 0.238 & 0.269 & 0.317 & 0.442\\
HAR (WLS, pooled) & 0.286 && 0.209 & 0.240 & 0.270 & 0.320 & 0.448\\
lasso & 0.290 && 0.211 & 0.240 & 0.273 & 0.327 & 0.434\\
Random Forest & 0.321 && 0.215 & 0.245 & 0.278 & 0.327 & 0.471\\
Gradient Boosted Trees & 0.333 && 0.217 & 0.250 & 0.290 & 0.335 & 0.499\\
Feedforward Neural Network & 0.369 && 0.233 & 0.258 & 0.302 & 0.415 & 0.680\\[0.5em]
    \textbf{VIX} \\
        \midrule
    HAR (OLS) & 0.275 && \textbf{0.201} & 0.226 & 0.257 & 0.307 & \textbf{0.416}\\
HAR (WLS) & \textbf{0.273} && 0.201 & \textbf{0.223} & \textbf{0.255} & \textbf{0.303} & 0.425\\
HAR (OLS, pooled) & 0.281 && 0.209 & 0.234 & 0.265 & 0.312 & 0.440\\
HAR (WLS, pooled) & 0.281 && 0.208 & 0.234 & 0.264 & 0.313 & 0.445\\
lasso & 0.285 && 0.203 & 0.230 & 0.263 & 0.306 & 0.460\\
Random Forest & 0.322 && 0.214 & 0.240 & 0.280 & 0.329 & 0.490\\
Gradient Boosted Trees & 0.350 && 0.210 & 0.238 & 0.278 & 0.327 & 0.492\\
Feedforward Neural Network & 0.410 && 0.222 & 0.256 & 0.307 & 0.439 & 0.888\\
    \bottomrule
        \end{tabular}%}
        \caption{Descriptive statistics for the MSEs of the different models without and with VIX. The table shows MSEs calculated during the test period from January 2022 to November 2023, averaged on a per-stock basis (within the NASDAQ-100 constituents). The lowest value in each column is highlighted in bold.}
        \label{tbl:mse_nasdaq100}
        \end{center}
    \end{table}
\begin{table}[h!]
        \begin{center}
        % \resizebox{\columnwidth}{!}{
        \begin{tabular}{l ccccccc}
        \toprule
        & & & \multicolumn{5}{c}{\textbf{Quantiles}} \\
        \cmidrule(lr){4-8}
        \textbf{No VIX} & \textbf{Mean} & & \textbf{5\%} & \textbf{25\%} & \textbf{50\%} & \textbf{75\%} & \textbf{95\%} \\
        \midrule
    HAR (OLS) & 0.186 && 0.117 & 0.140 & 0.170 & 0.225 & 0.301\\
HAR (WLS) & \textbf{0.179} && \textbf{0.113} & \textbf{0.136} & \textbf{0.164} & \textbf{0.219} & \textbf{0.288}\\
HAR (OLS, pooled) & 0.189 && 0.119 & 0.142 & 0.170 & 0.233 & 0.308\\
HAR (WLS, pooled) & 0.183 && 0.114 & 0.138 & 0.164 & 0.227 & 0.297\\
lasso & 0.201 && 0.125 & 0.156 & 0.185 & 0.238 & 0.337\\
Random Forest & 0.212 && 0.125 & 0.159 & 0.192 & 0.244 & 0.332\\
Gradient Boosted Trees & 0.216 && 0.125 & 0.163 & 0.197 & 0.250 & 0.342\\
Feedforward Neural Network & 0.257 && 0.136 & 0.166 & 0.205 & 0.319 & 0.515\\[0.5em]
    \textbf{VIX} \\
        \midrule
    HAR (OLS) & 0.190 && 0.117 & 0.145 & 0.172 & 0.232 & 0.323\\
HAR (WLS) & 0.182 && 0.114 & 0.139 & 0.165 & 0.221 & 0.303\\
HAR (OLS, pooled) & 0.192 && 0.121 & 0.144 & 0.172 & 0.236 & 0.315\\
HAR (WLS, pooled) & 0.185 && 0.117 & 0.137 & 0.165 & 0.229 & 0.302\\
lasso & 0.194 && 0.115 & 0.148 & 0.175 & 0.224 & 0.309\\
Random Forest & 0.213 && 0.120 & 0.153 & 0.189 & 0.240 & 0.326\\
Gradient Boosted Trees & 0.216 && 0.117 & 0.153 & 0.186 & 0.240 & 0.328\\
Feedforward Neural Network & 0.290 && 0.123 & 0.165 & 0.196 & 0.283 & 0.813\\
    \bottomrule
        \end{tabular}%}
        \caption{Descriptive statistics for the QLIKE of the different models without and with VIX. The table shows QLIKE calculated during the test period from January 2022 to November 2023, averaged on a per-stock basis (within the NASDAQ-100 constituents). The lowest value in each column is highlighted in bold.}
        \label{tbl:qlike_nasdaq100}
        \end{center}
    \end{table}
\begin{table}[h!]
    \begin{center}
    \resizebox{\columnwidth}{!}{
    \begin{tabular}{l ccccccc}
    \toprule
    & & & \multicolumn{5}{c}{\textbf{Quantiles}} \\
    \cmidrule(lr){4-8}
    \textbf{No VIX} & \textbf{Mean} & & \textbf{5\%} & \textbf{25\%} & \textbf{50\%} & \textbf{75\%} & \textbf{95\%} \\
    \midrule
HAR (OLS) & 3.573\% && 3.267\% & 3.467\% & 3.622\% & 3.687\% & 3.748\%\\
HAR (WLS) & \textbf{3.594}\% && \textbf{3.309}\% & \textbf{3.489}\% & \textbf{3.639}\% & 3.700\% & 3.758\%\\
HAR (OLS, pooled) & 3.564\% && 3.255\% & 3.452\% & 3.614\% & 3.689\% & 3.748\%\\
HAR (WLS, pooled) & 3.584\% && 3.289\% & 3.475\% & 3.634\% & \textbf{3.703}\% & \textbf{3.761}\%\\
lasso & 3.530\% && 3.171\% & 3.454\% & 3.578\% & 3.650\% & 3.728\%\\
Random Forest & 3.514\% && 3.228\% & 3.428\% & 3.548\% & 3.634\% & 3.726\%\\
Gradient Boosted Trees & 3.507\% && 3.209\% & 3.447\% & 3.550\% & 3.629\% & 3.729\%\\
Feedforward Neural Network & 3.398\% && 2.728\% & 3.306\% & 3.531\% & 3.634\% & 3.709\%\\[0.5em]
    \textbf{VIX} \\
    \midrule
HAR (OLS) & 3.554\% && 3.228\% & 3.454\% & 3.600\% & 3.669\% & 3.746\%\\
HAR (WLS) & 3.578\% && 3.282\% & 3.477\% & 3.619\% & 3.686\% & 3.755\%\\
HAR (OLS, pooled) & 3.553\% && 3.227\% & 3.455\% & 3.605\% & 3.678\% & 3.740\%\\
HAR (WLS, pooled) & 3.575\% && 3.268\% & 3.478\% & 3.627\% & 3.694\% & 3.755\%\\
lasso & 3.549\% && 3.279\% & 3.486\% & 3.592\% & 3.669\% & 3.750\%\\
Random Forest & 3.510\% && 3.265\% & 3.441\% & 3.561\% & 3.654\% & 3.742\%\\
Gradient Boosted Trees & 3.515\% && 3.244\% & 3.445\% & 3.574\% & 3.657\% & 3.745\%\\
Feedforward Neural Network & 3.309\% && 1.855\% & 3.342\% & 3.545\% & 3.633\% & 3.741\%\\
    \bottomrule
    \end{tabular}}
    \caption{Descriptive statistics for the realized utilities (without transaction costs) of the different models without and with VIX. The table shows realized utilities calculated during the test period from January 2022 to November 2023 for the NASDAQ-100 constituents. The largest value in each column is highlighted in bold.}
    \label{tbl:ru_nasdaq100}
    \end{center}
\end{table}
\begin{table}[h!]
    \begin{center}
    \resizebox{\columnwidth}{!}{
    \begin{tabular}{l ccccccc}
    \toprule
    & & & \multicolumn{5}{c}{\textbf{Quantiles}} \\
    \cmidrule(lr){4-8}
    \textbf{No VIX} & \textbf{Mean} & & \textbf{5\%} & \textbf{25\%} & \textbf{50\%} & \textbf{75\%} & \textbf{95\%} \\
    \midrule
HAR (OLS) & 2.666\% && -0.279\% & 2.419\% & 3.215\% & 3.533\% & 3.652\%\\
HAR (WLS) & \textbf{2.688}\% && \textbf{-0.239}\% & \textbf{2.450}\% & \textbf{3.237}\% & \textbf{3.545}\% & 3.663\%\\
HAR (OLS, pooled) & 2.658\% && -0.273\% & 2.426\% & 3.207\% & 3.520\% & 3.654\%\\
HAR (WLS, pooled) & 2.678\% && -0.246\% & 2.447\% & 3.226\% & 3.535\% & 3.665\%\\
lasso & 2.623\% && -0.365\% & 2.298\% & 3.161\% & 3.482\% & 3.627\%\\
Random Forest & 2.607\% && -0.394\% & 2.298\% & 3.172\% & 3.458\% & 3.625\%\\
Gradient Boosted Trees & 2.600\% && -0.385\% & 2.224\% & 3.164\% & 3.463\% & 3.616\%\\
Feedforward Neural Network & 2.491\% && -0.340\% & 2.145\% & 3.040\% & 3.425\% & 3.625\%\\[0.5em]
    \textbf{VIX} \\
    \midrule
HAR (OLS) & 2.648\% && -0.343\% & 2.330\% & 3.205\% & 3.507\% & 3.650\%\\
HAR (WLS) & 2.672\% && -0.295\% & 2.368\% & 3.222\% & 3.534\% & 3.659\%\\
HAR (OLS, pooled) & 2.646\% && -0.282\% & 2.414\% & 3.200\% & 3.514\% & 3.644\%\\
HAR (WLS, pooled) & 2.668\% && -0.254\% & 2.439\% & 3.223\% & 3.531\% & \textbf{3.665}\%\\
lasso & 2.642\% && -0.322\% & 2.334\% & 3.211\% & 3.502\% & 3.649\%\\
Random Forest & 2.603\% && -0.382\% & 2.241\% & 3.204\% & 3.465\% & 3.643\%\\
Gradient Boosted Trees & 2.609\% && -0.382\% & 2.211\% & 3.202\% & 3.470\% & 3.636\%\\
Feedforward Neural Network & 2.402\% && -0.767\% & 2.094\% & 3.140\% & 3.480\% & 3.624\%\\
    \bottomrule
    \end{tabular}}
    \caption{Descriptive statistics for the realized utilities (with transaction costs) of the different models without and with VIX. The table shows realized utilities calculated during the test period from January 2022 to November 2023 for the NASDAQ-100 constituents. The largest value in each column is highlighted in bold.}
    \label{tbl:rutc_nasdaq100}
    \end{center}
\end{table}

\begin{figure}[!htb]
    \centering
    \includegraphics[width=\textwidth]{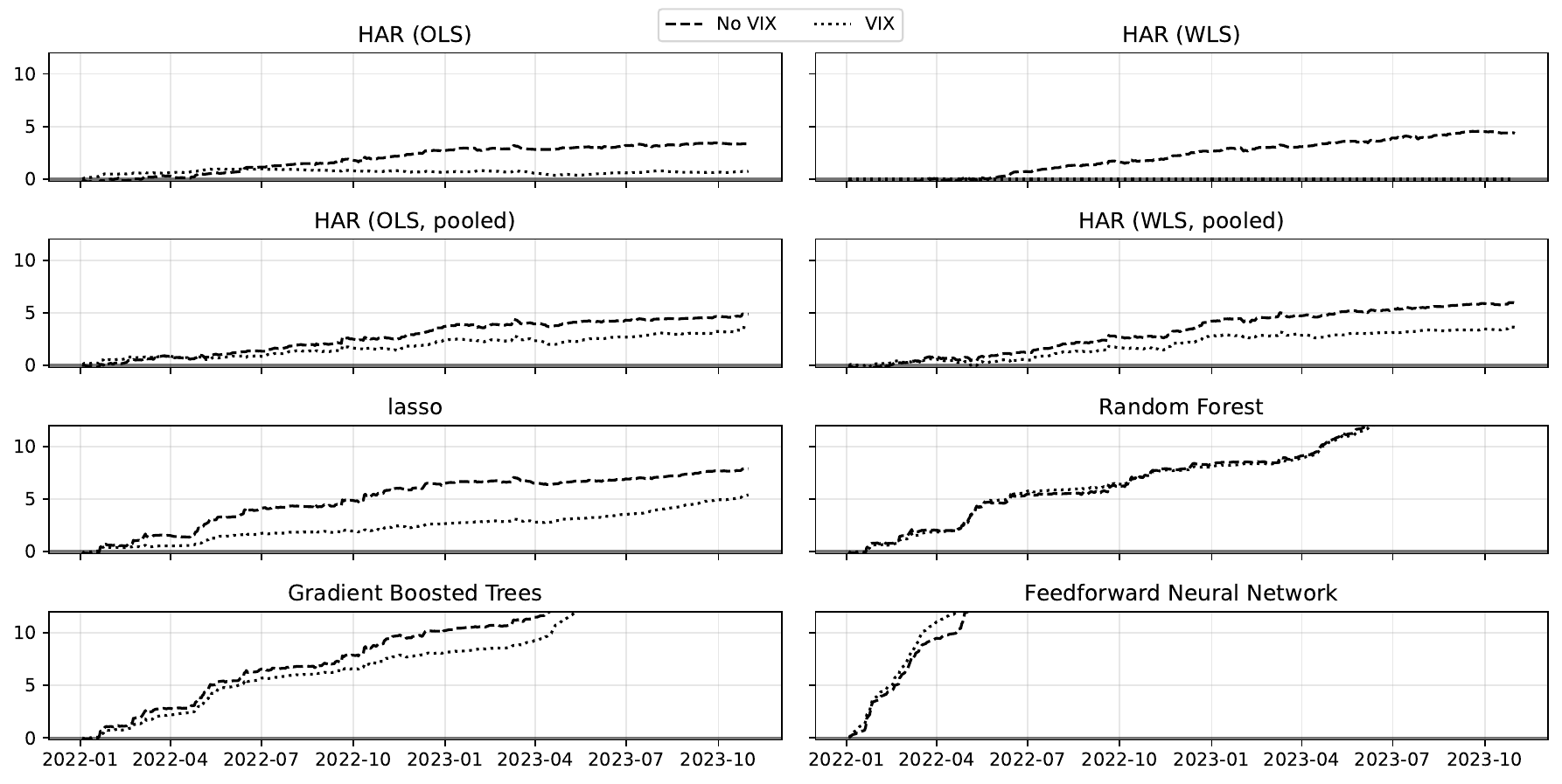}
    \caption{Cumulative squared error difference between the models and the HAR-VIX model estimated via WLS for the Nasdaq-100 constituents. The dashed lines represent the models without the VIX, while the dotted lines represent the models when the VIX is included. For better visualization, the $y$-axis is constrained to the range $[0, 12]$.}
    \label{fig:cse_diff_nasdaq100}
\end{figure}
\setcounter{table}{0}
\clearpage
\renewcommand{\thetable}{D.\arabic{table}}
\section{Hyperparameter Tuning}\label{sec:hyperparameters}

Hyperparameter tuning is a crucial step in the development of predictive ML models, as it involves selecting the set of parameters that are not directly learned during the training process.
Nonetheless, the optimal hyperparameters can vary significantly across different datasets and models, making it a challenging task to find the best configuration.
While proper tuning of hyperparameters can drastically influence the performance of ML models, it is a computationally expensive process that hinders the practical application of such models in financial forecasting, where real-time predictions are often required and rolling windows are commonly used.

\begin{table}[h!]
    \begin{center}
    % \resizebox{\columnwidth}{!}{
    \begin{tabular}{l cc}
    \toprule
     & \textbf{Hyperparameter} & \textbf{Candidate Values} \\
    \cmidrule(lr){2-3}
    % \textbf{HAR} & Training window & $\{5 + 10i\}_{i=0}^{200}$\\
    % & Stride & $\{i\}_{i=1}^{10} \cup \{10 + 3i\}_{i=1}^{170}$\\
    % \midrule
    \textbf{Lasso} & $\lambda$ & $\{10^{2-(5i/1000)}\}_{i=0}^{1000}$\\[0.5em]

    \textbf{Random Forest} & Num. of trees & $100, 250, 500$\\
    & Min. samples per leaf & $1, 5, 10$\\
    & Max. features & $p / 3, \sqrt{p}, p$\\[0.5em]

    \textbf{Gradient Boosted Trees} & Depth & $1,2$\\
    & Num. of trees & $100, 250, 500$\\
    & Learning rate & $0.01, 0.1$\\[0.5em]

    \textbf{Feedforward Neural Network} & $\lambda$ & $0, 10^{-4}$\\ 
    & Architecture & $(2,), (4,2), (8,4,2), $\\
    & & $(16,8,4,2), (32,32), (64,64)$\\
    \bottomrule
    \end{tabular}%}
    \caption{Candidate values for the hyperparameters of the different models}
    \label{tbl:hyperparameters}
    \end{center}
\end{table}

In Table~\ref{tbl:hyperparameters}, we outline the candidate values for the hyperparameters of our ML models.
The selection of these candidate values is informed by prior research conducted by \citet{Zhang2023,Christensen2023}.
We ensure that our hyperparameter search space is expansive enough to cover all configurations explored in these studies. Additionally, we have included other values that we anticipate may be beneficial, while maintaining a search space that remains computationally manageable given our large dataset.
This approach is focused on striking a balance between allowing for comparability with previous studies, ensuring that we can validate our results against the existing literature, exploring new configurations that may improve the performance of our models to give our ML models the best chance of success, and maintaining computational feasibility.

We give a brief description of the hyperparameters for each model below:
\begin{itemize}\setlength\itemsep{0em}
    \item \textbf{lasso}: The hyperparameter $\lambda$ controls the strength of the regularization applied to the model, with higher values increasing the penalty imposed on the magnitude of the coefficients, as per \eqref{eq:lasso}.
    \item \textbf{Random Forest}: \textit{Num. of trees} dictates how many trees are built in the ensemble, with a higher number typically leading to better performance. \textit{Min. samples per leaf} defines the smallest number of samples required to be at a leaf node, which can prevent overfitting. \textit{Max. features} determines the number of features when looking for the best split, with a lower number potentially decreasing the correlation between trees and improving performance.
    \item \textbf{Gradient Boosted Trees}: \textit{Depth} specifies the maximum depth of the individual trees, with deeper trees potentially capturing more complex patterns. \textit{Num. of trees} and \textit{learning rate} control the number of trees in the ensemble and the contribution of each tree, respectively.
    \item \textbf{Feedforward Neural Network}: The hyperparameter $\lambda$ is the regularization term in the loss function, which penalizes the squared Euclidean norm of the coefficients, similar to Ridge regression \citep{Hoerl1970}. \textit{Architecture} describes the structure of the network, with the position of the numbers indicating the layers and the value of each number representing the number of neurons in that layer; e.g., $(8, 4, 2)$ denotes a network with three layers, with eight neurons in the first layer, four in the second, and two in the third. Note that $(2,)$ denotes a single-layer network with two neurons.
\end{itemize}

% \section*{Acknowledgements}
% Thank people.
% List funding agencies.
% Comment this section out if it reveals the authors' identities.

\end{document}